\documentclass[aps,pre,twocolumn,showpacs,superscriptaddress,showkeys,showpacs,groupedaddress]{revtex4}
\usepackage{graphics}
\usepackage{amsmath}
\usepackage{amssymb}
\usepackage{amsfonts}
\usepackage{graphicx}
\usepackage{color}
\usepackage{soul}
\usepackage[caption=false]{subfig}

\graphicspath{{paper_plots/}{alex_plots/}{./}{philipp_plots/}}

\begin{document}
\title{Chimera patterns in two-dimensional networks of coupled neurons}
% \title{Chimera patterns in two-dimensional networks of FitzHugh-Nagumo and Leaky Integrate-and-Fire oscillators}
\author{Alexander Schmidt} 
\affiliation{Institut f{\"u}r Theoretische Physik, Technische Universit{\"a}t Berlin, Hardenbergstra\ss{}e 36, 10623 Berlin, Germany}
\author{Theodoros Kasimatis}
\affiliation{Institute of Nanoscience and Nanotechnology, National Center for Scientific Research ``Demokritos'', 15310 Athens, Greece}
\author{Johanne Hizanidis}
\affiliation{Institute of Nanoscience and Nanotechnology, National Center for Scientific Research ``Demokritos'', 15310 Athens, Greece}
\affiliation{Crete Center for Quantum Complexity and Nanotechnology, Department of Physics, University of Crete, 71003 Heraklion, Greece}
\author{Astero Provata}
\affiliation{Institute of Nanoscience and Nanotechnology, National Center for Scientific Research ``Demokritos'', 15310 Athens, Greece}
\author{Philipp H{\"o}vel} 
% \email{phoevel@physik.tu-berlin.de}
\affiliation{Institut f{\"u}r Theoretische Physik, Technische Universit{\"a}t Berlin, Hardenbergstra\ss{}e 36, 10623 Berlin, Germany}
\affiliation{Bernstein Center for Computational Neuroscience Berlin, Humboldt-Universit{\"a}t zu Berlin,
Philippstra{\ss}e 13, 10115 Berlin, Germany}

\date{Received: \today / Revised version: date}
% The correct dates will be entered by Springer
%
\begin{abstract}{
We discuss synchronization patterns in networks of FitzHugh-Nagumo 
and Leaky Integrate-and-Fire oscillators coupled in a two-dimensional toroidal 
geometry. Common feature between the two models is the presence of fast and slow dynamics, a typical characteristic of neurons.
Earlier studies have demonstrated that both models when coupled nonlocally in one-dimensional ring 
networks produce chimera states for a large range of parameter values.
In this study, we give evidence of a plethora of two-dimensional chimera patterns of various shapes including spots, rings, 
stripes, and grids, observed in both models, as well as additional patterns found mainly in the FitzHugh-Nagumo system.
Both systems exhibit multistability: For the same parameter values, different initial conditions give rise to
different dynamical states. Transitions occur between various patterns when the parameters (coupling range,
coupling strength, refractory period, and coupling phase) are varied. 
Many patterns observe in the two models follow similar rules. For example the diameter of the rings grows linearly with the coupling radius.
}
\end{abstract}

 % end of PACS codes
%      %end of abstract
%
\maketitle
\section{Introduction}
\label{sec:intro}
Chimeras are hybrid states that emerge spontaneously, combining both coherent and incoherent parts~\cite{abrams:2004}. First
found in identical and symmetrically coupled phase oscillators~\cite{kuramoto:2002}, chimera states have 
been the focus of extensive research for over a decade now. Both theoretical and experimental works have
shown that this counter-intuitive collective phenomenon may arise in numerous systems including mechanical, chemical, electro-chemical,
electro-optical, electronic, and superconducting coupled oscillators~\cite{panaggio:2015,tinsley:2012,hagerstrom:2012,wickramasinghe:2013,martens:2013,
Rosin2014,schmidt:2014,Gambuzza2014,Kapitaniak2014,LAZ15,HIZ16b,omelchenko:2011}. 

The phenomenon of chimera states has also been addressed in networks of biological neural oscillators.
In particular, Hindmarsh-Rose neural oscillators have been studied in networks with nonlocal~\cite{HIZ13} and nearest-neighbor~\cite{BER16}
coupling, as well as in modular networks consisting of communities~\cite{HIZ16}.
Potential relevance of chimera states in this context include bump states~\cite{LAI01,SAK06a} and the phenomenon of unihemispheric sleep
observed in birds and dolphins~\cite{RAT00}, which sleep with one eye open, meaning that half of the brain is synchronous with the other half being asynchronous.
Furthermore, it has been recently hypothesized that chimera states are the route of onset or termination of epileptic
seizures~\cite{MOR00,MOR03a,ROT14,AND16}.

Chimera states have also been reported in the one-dimensional FitzHugh-Nagumo (FHN) model with nonlocal
connectivity~\cite{omelchenko:2013}. Patches of synchrony were observed
within the incoherent domains giving rise to multi-chimera states,
when the coupling constant increased above the weak limit. 
The multiplicity of the state, that is, the number of (in)coherent regions, depended on the coupling strength and range.
These multi-chimera states were shown to be robust when small
inhomogeneities in the coupling topology (with identical elements)
or inhomogeneous elements with regular nonlocal coupling were introduced~\cite{omelchenko:2015}. 
For a constant coupling strength and given number of links, hierarchical connectivity
 was shown to induce nested multi-chimera patterns~\cite{omelchenko:2015}.
Coexistence of coherent and incoherent domains were also observed in systems of excitable FHN elements under the influence of noise.
As Semenova \emph{et al.} stress, the noise has often a constructive role: It
shifts the dynamics of identical excitable elements into the oscillatory
domain, giving then rise to chimera states~\cite{semenova:2016}. 
Moreover, it is possible to control the 
position of coherent and incoherent domains of a multi-chimera state
by introducing a block of excitable FHN elements in appropriate positions,
or to generate a chimera directly from the synchronous state~\cite{isele:2016}. 
This observation offers
promising ideas in terms of achieving desired states
by a local modification of the system parameters.
For instance, it may be possible in targeted medication to modify incoherent neuron dynamics by changing
only the potential of a few neurons locally, leaving the coupling topology of the rest of the network untouched.
In addition, Omelchenko \emph{et al.} proposed an alternative control scheme that extends the lifetime of chimeras, which are known to be chaotic transients~\cite{WOL11}, and at the same time reduces the erratic drift present in small networks~\cite{OME16}. Controlling the position can also be achieved by introducing an asymmetric coupling strength~\cite{BIC15}.

Experimentally, the FHN model has been studied by Essaki Arumugam and Spano, in connection to
synchronization phenomena associated with neurological disorders such as epilepsy~\cite{essaki:2015}.
In their study, they implemented nine FHN neurons linked in a ring topology,
via discrete electronics. Introducing nonlocal coupling, a chimera state appeared, while for local connectivity
either fully synchronous or asynchronous states were observed. Their results indicate that 
epilepsy can be understood as a topological disease, strongly related to 
the connectivity of the underlying network of neurons. 

Chimera states were recently reported in the one-dimensional Leaky Integrate-and-Fire (LIF) model, a system describing
the spiking behavior of neuron cells. It was shown that identical LIF oscillators nonlocally linked in
a one-dimensional ring geometry give rise to multi-chimera states whose multiplicity depends
both on the coupling strength and the refractory period of the neuron cells~\cite{tsigkri:2016}.
In analogy with the FHN model,
the introduction of a hierarchical topology in the coupling induced
nested chimera states and also transitions between multi-chimera states with different multiplicities. 
Using a different geometrical setup of two populations of identical LIF 
oscillators coupled via excitatory coupling, Olmi \emph{et al.} studied the onset of chimeras 
as well as states characterized by a different degree of synchronization in the two populations~\cite{olmi:2010}. Irregular synchronization phenomena have been reported for LIF
elements even in the case of all-to-all connectivity~\cite{luccioli:2010}.

Chimera states have mainly been investigated in one-dimensional systems.
Recently, works involving two~\cite{OME12,XI15}
and three-dimensional~\cite{MAI15} oscillator arrays 
have revealed new types of chimera states, depending on the 
coupling function. These studies have mainly focused on phase oscillators.
In this paper, we go beyond the simple Kuramoto model and consider
a two-dimensional network configuration using two different models related to neuronal spiking activity.
This topology is motivated from medical experiments where thin brain slices are cultured in Petri dishes and various electrical and chemical properties are recorded, see for example Ref.~\cite{meijer:2015}. The two-dimensional nonlocal connectivity studied here can be considered as
 an approximation of the acute brain slices, whose connectivity is certainly more complex.

Our investigation follows a parallel presentation
of common patterns in the two-dimensional FHN and LIF systems highlighting the main features that
are responsible for the formation of each pattern in the two systems. In Sec.~\ref{sec:models}, the two models are briefly recapitulated. In Sec.~\ref{sec:spots},
main attributes of spot and ring chimeras are presented for the two models, and their 
common and different properties are discussed. Similarly, in Secs.~\ref{sec:stripes} and~\ref{sec:grid} stripe and grid chimeras are discussed, respectively. Additional miscellaneous chimeras as well as other patterns are summarized in the appendices. Finally, the main results and open problems
are discussed in a brief concluding section.\\

% \vskip 1cm
\section{The Models}
\label{sec:models}
In this section, we summarize the two models considered in this article -- FitzHugh-Nagumo (FHN) and Leaky Integrate-and-Fire (LIF) model -- and introduce the respective toroidal coupling schemes describing the two-dimensional layout with nonlocal interactions.

\subsection{The FitzHugh-Nagumo model}

\label{sec:FHN}
The dynamics of a single FHN system are described by the following set of equations~\cite{FIT61,NAG62}:
\begin{subequations}
\begin{align}
\epsilon \frac{dx}{dt}&=x-\frac{x^3}{3}-y\\
\frac{dy}{dt}&=x+a,
\end{align}
\end{subequations}
where $x$ and $y$ denote a fast activator and slow inhibitor variable, respectively. 
The parameter $\epsilon$ determines the timescale separation and is fixed in this study at $\epsilon=0.05$. The threshold parameter $a$ determines the oscillatory ($|a| <1$) or excitable ($|a|>1$) behavior in the system, i.e., an unstable or stable fixed point at the intersection of the nullclines, and is set to $a=0.5$ throughout this study. The interaction of the two system variables of each node in the network is realized via a rotational coupling matrix as proposed in Ref.~\cite{omelchenko:2013} and detailed below.

\begin{figure}[ht!]
\centering
 \includegraphics[width=1\linewidth]{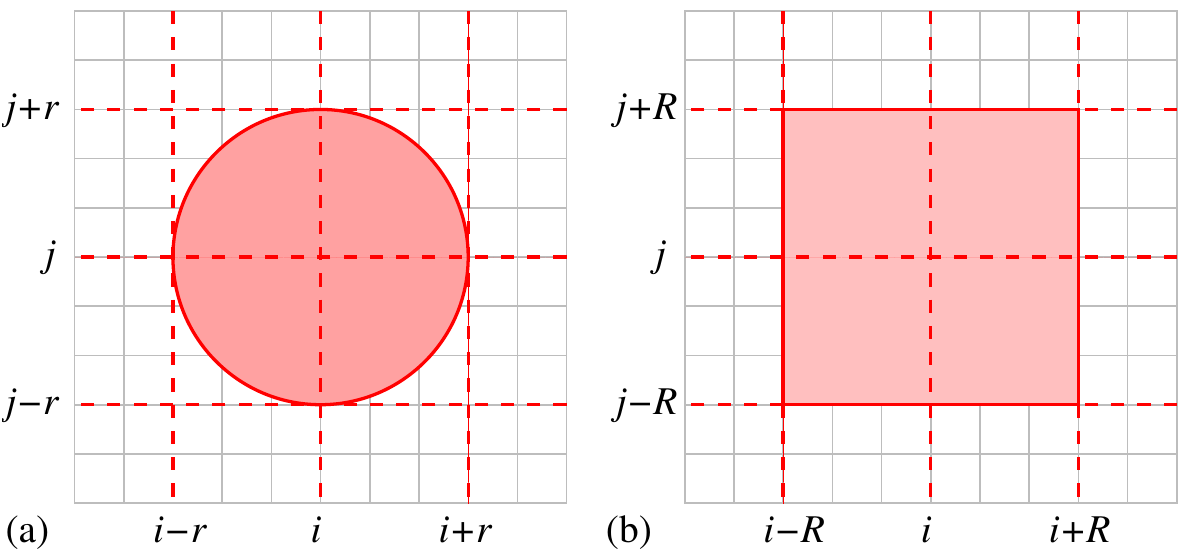}
\caption{(Color online) Schematic depiction of the coupling topology used for the FHN model of Eqs.~\eqref{eq:fitzhugh} in panel (a) and for the LIF model of Eq.~\eqref{eq:LIF_07} in panel (b) according to Eqs.~\eqref{eq:ball} and~\eqref{eq:LIF_08}, respectively.}
\label{fig:topology}
\end{figure}

We consider the following coupling scheme for the FHN model: A two-dimensional regular $N \times N$-network with $N=100$ nodes and periodic, toroidal boundary conditions. This setting yields the following system of network equations:
\begin{widetext}
\begin{subequations}
\begin{align}
\epsilon \frac{dx_{ij}}{dt}&=x_{ij}-\frac{x_{ij}^3}{3}-y_{ij}
+\frac{\sigma}{N_r-1} \sum\limits_{(m,n) \in B_r (i,j)}[b_{xx}(x_{ij}-x_{mn})+b_{xy}(y_{ij}-y_{mn})]\\
\frac{dy_{ij}}{dt}&=x_{ij}+a+\frac{\sigma}{N_r-1} \sum\limits_{(m,n) \in B^{\text{FHN}}_r (i,j)}[b_{yx}(x_{ij}-x_{mn})+b_{yy}(y_{ij}-y_{mn})],
\end{align}
\label{eq:fitzhugh}
\end{subequations}
\end{widetext}
where we fix the coupling strength at $\sigma=0.1$ in the following.
Note that the double index of the dynamic variables $x_{ij}$ and $y_{ij}$ with $i,j=1, \dots, N$ refers to the position
on the two-dimensional lattice. All oscillators $(x_{ij},y_{ij})$ are identical and 
couple isotropically to all other oscillators in a circular neighborhood given by:
\begin{equation}
B^{\text{FHN}}_r (i,j)=\{ (m,n):(m-i)^2+(n-j)^2 \leq r^2 \},
\label{eq:ball}
\end{equation}
where the differences need to be calculated in a manner to account for a toroidal geometry. In other words, $(m-i)^2+(n-j)^2$ is computed to obtain the shortest distance on the torus. See Fig.~\ref{fig:topology}(a) for a schematic depiction.
Hence, the number of lattice points inside a circle of radius $r$ is given by:
% (Gauss's circle problem\footnote{Weisstein, Eric W. ''Gauss's Circle Problem.'' From \textit{MathWorld} - A Wolfram Web Resource. \url{http://mathworld.wolfram.com/GausssCircleProblem.html}, accessed 2016-04-23}) can be calculated with:
%
\begin{equation}
N_r = 1+4 \sum\limits_{i=0}^\infty \left ( \left \lfloor \frac{r^2}{4i+1} \right \rfloor- \left \lfloor \frac{r^2}{4i+3} \right \rfloor \right ),
\label{eq:gauss}
\end{equation}
where $\lfloor$\textperiodcentered$\rfloor$ is the floor function, which gives the largest integer less than or equal to its argument.
% \footnote{Weisstein, Eric W. ''Floor Function.'' From \textit{MathWorld} - A Wolfram Web Resource. \url{http://mathworld.wolfram.com/FloorFunction.html}, accessed 2016-04-23}. 

The coupling between the $x$- and $y$-variables is realized by a rotational coupling matrix using the single parameter $\varphi \in [0,2\pi)$, which allows diagonal or direct coupling ($\varphi=0$ and $\varphi=\pi$), cross coupling ($\varphi=\pi/2$ and $\varphi =3 \pi /2$) and mixed coupling scenarios. Accordingly, the coupling matrix is given by:
\begin{equation}
B=
\begin{pmatrix}
b_{xx} & b_{xy}\\
b_{yx} & b_{yy}
\end{pmatrix}
=\begin{pmatrix}
\cos{\varphi} & \sin{\varphi}\\
-\sin{\varphi} & \cos{\varphi}
\end{pmatrix}.
\end{equation}
For this type of coupling, the coupling phase $\varphi$ corresponds to the so-called phase frustration parameter $\alpha$ in the Kuramoto system, which is known to be a crucial quantity for the occurrence of chimera states~\cite{OME10a}.
% \red{cite Omel'chenko Phys. Rev.~E 81, 065201(R) (2010) and Phys. Rev.~E 85, 036210 (2012)}. 
As derived in Ref.~\cite{omelchenko:2013}, the mapping from $\varphi$ to $\alpha$ is given via a phase-reduction technique.

In the following, we will use the coupling phase $\varphi$ and coupling radius $r$ to explore the dynamical scenarios in the networked system.

\subsection{The Leaky Integrate-and-Fire model}
\label{sec:LIF}
The LIF model describes a neuron
% \cite{ladenbauer:2013,brunel:2007,kouvaris:2010,lucioli:2010}
via a single dynamical variable $u$ that can be interpreted as its membrane potential. 
The variable $u$ evolves according to the following equation:
\begin{equation}\label{eq:LIF_01}
\frac{du}{dt}=\mu-u.
\end{equation}
If the membrane potential $u$ reaches a threshold $u_\text{th}<\mu$, it is reset to a resting potential:
\begin{equation}
\lim_{\varepsilon \to 0} u(t+\varepsilon ) \to u_\text{rest}, 
\>\>\> \text{when} \>\> u\ge u_\text{th}.
\label{eq:LIF_02}
\end{equation}
Without loss of generality, the reset or resting potential can be set to $u_\text{rest} = 0$. 

The solution of Eq.~\eqref{eq:LIF_01} is then given by
$u(t)=\mu -\left(\mu-u_0\right)e^{-t}$
% \begin{equation}
% u(t)=\mu -\left(\mu-u_0\right)e^{-t}
% \label{eq:LIF_03} 
% \end{equation}
with the initial condition $u(0)=u_0$ and is repeated after each reset starting again at $u_0=u_{\text{rest}}$. The period between two resets $T_s$ of 
a single element depends
both on $\mu$ and $u_\text{th}$ as follows:
$T_s=\ln \left[\left(\mu -u_0\right)/\left(\mu - u_\text{th}\right)\right]$.
% \begin{equation}
% T_s=\ln \frac{\mu -u_0}{\mu - u_\text{th}}.
% \label{eq:LIF_04}
% \end{equation}

As an extension to the standard LIF model, we take a refractory period $p_r$ into account via the following condition:
\begin{align}
u(t)=u_{\rm rest}, \forall t: [n (T_s+p_r)+T_s] \le t \le [(n+1)(T_s+p_r)] 
\label{eq:LIF_05}
\end{align}
with $n=0,1,2,\dots$. In other words, the LIF element is held at the rest potential during the refractory period. As a result, the 
total period $T$ of the single element is now:
$T=T_s+p_r=\ln\left[\left(\mu -u_0\right)/\left(\mu - u_\text{th}\right)\right]+p_r$.
% \begin{equation}
% T=T_s+p_r=\ln \frac{\mu -u_0}{\mu - u_\text{th}}+p_r.
% \label{eq:LIF_06} 
% \end{equation}

Similar to the case of the FHN model, we consider the LIF network dynamics on
a two-dimensional regular $N \times N$-network with $N=100$ nodes and toroidal boundary conditions. The LIF coupled network dynamics
read:
\begin{equation}
\label{eq:LIF_07} 
\frac{du_{ij}}{dt}=\mu-u_{ij}+\frac{\sigma}{N_R-1} \sum_{(m,n)\in B^{\text{LIF}}_R (i,j) } \left[u_{ij} - u_{mn}\right],
\end{equation}
where the neighborhood of the element $u_{ij}$ with $i,j=1,\dots,N$ is given by
\begin{equation}
B^{\text{LIF}}_R (i,j)=\{ (m,n):i-R \leq m \le i+R \,\wedge\, j-R \leq n \le j+R \},
\label{eq:LIF_08}
\end{equation}
which includes all elements on the grid within a square of side $2R+1$ as depicted in Fig.~\ref{fig:topology}(b). The size of the
coupled region is then given by $N_R=(2R+1)^{2}$.

To calculate coherence in both models we use the mean phase velocity $\omega_{ij}$ of an oscillator at position $(i,j)$
\cite{omelchenko:2013}.
If $c_{ij}(\Delta t)$ is the number of periods that the oscillator at position $(i,j)$ has completed in a time interval $\Delta t$, the mean phase velocity is
defined as:
$\omega_{ij}=2\pi c_{ij}(\Delta t)/\Delta t$.
% \begin{equation}
% \label{eq:LIF_09}
% \omega_{ij}={{2\pi c_{ij}(t)}\over{t}}.
% \end{equation}

In the next sections, we provide a series of selected patterns that we observe for both the FHN and LIF model. 
In order to explore the parameter space, we use random initial conditions. Whenever we observe a characteristic pattern for a specific choice of coupling parameters, we perform a continuation by scanning the vicinity using the pattern of the previous simulation as an initial condition. This way, we are able to map out the regions of existence of the different dynamical patterns.

\section{Spots and Ring Patterns}
\label{sec:spots}
In this section, we focus on chimera states of spot and ring type in both FHN and LIF models.
\subsection{FitzHugh-Nagumo: Spots and ring patterns}
\label{sec:spots_FHN}
In the $100\times100$ FHN system described by Eqs.~\eqref{eq:fitzhugh} and~\eqref{eq:ball}, we observe various chimera regions in the parameter space spanned by the coupling radius $r\in[1,49]$ and coupling phase $\varphi\in[0,2\pi)$. A complete scan of the parameter space and several enlargements can be found in Appendix~\ref{sec:fhn-parameter-scan}. Here, we will focus on the range $\varphi\in[1.3,1.65]$. 

\begin{figure}[ht!]
\includegraphics[width=1\linewidth,angle=0]{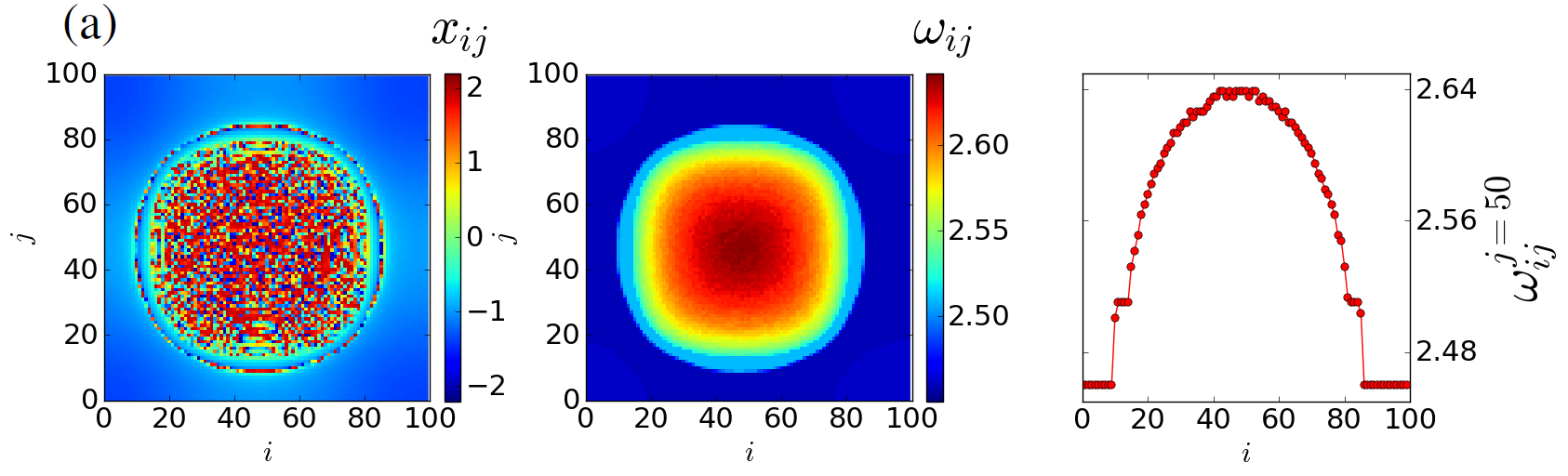}
\includegraphics[width=1\linewidth]{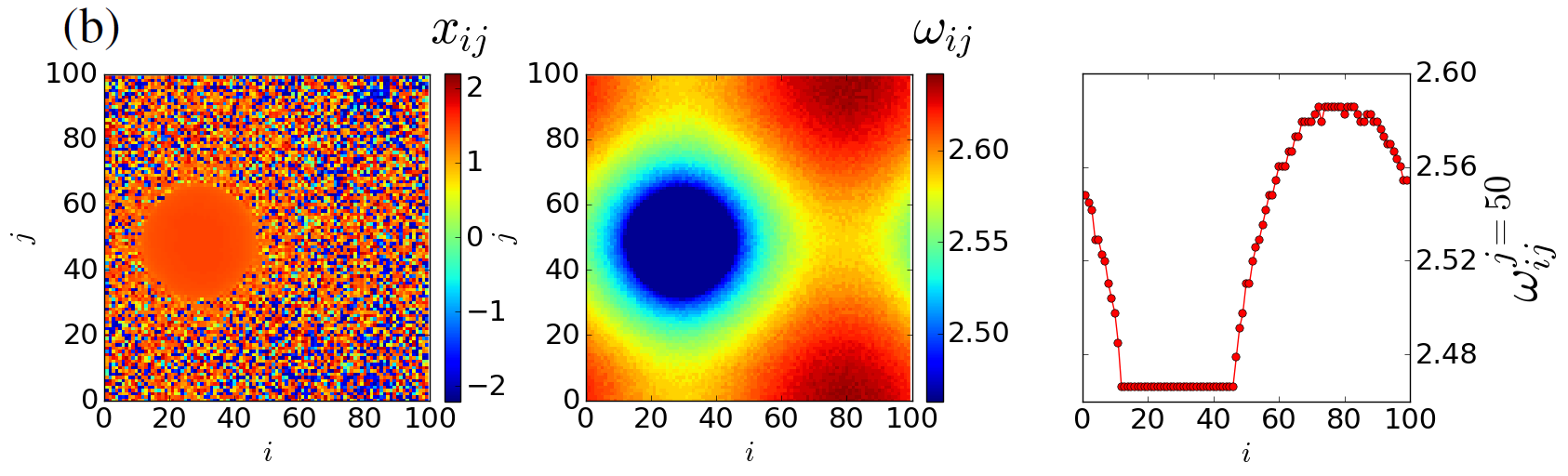}
\caption{(Color online) FHN system: 
Chimera state of (a) incoherent-spot and (b) coherent-spot type. The left panels depict a snapshot of the activator variable~$x_{ij}$ at $t=2000$. The center and right panels show the mean phase velocity $\omega_{ij}$ and a horizontal cut of $\omega_{ij}$. Coupling parameters: (a) $r = 33$, $\varphi=\pi/2 -0.2$ and (b) $r=42$, $\varphi=\pi/2$. Other parameters: $\sigma=0.1$, $a=0.5$, and $\epsilon=0.05$.}
\label{fig:fhnSpots1}
\end{figure}

\begin{figure}[ht!]
\centering
 \includegraphics[width=1\linewidth]{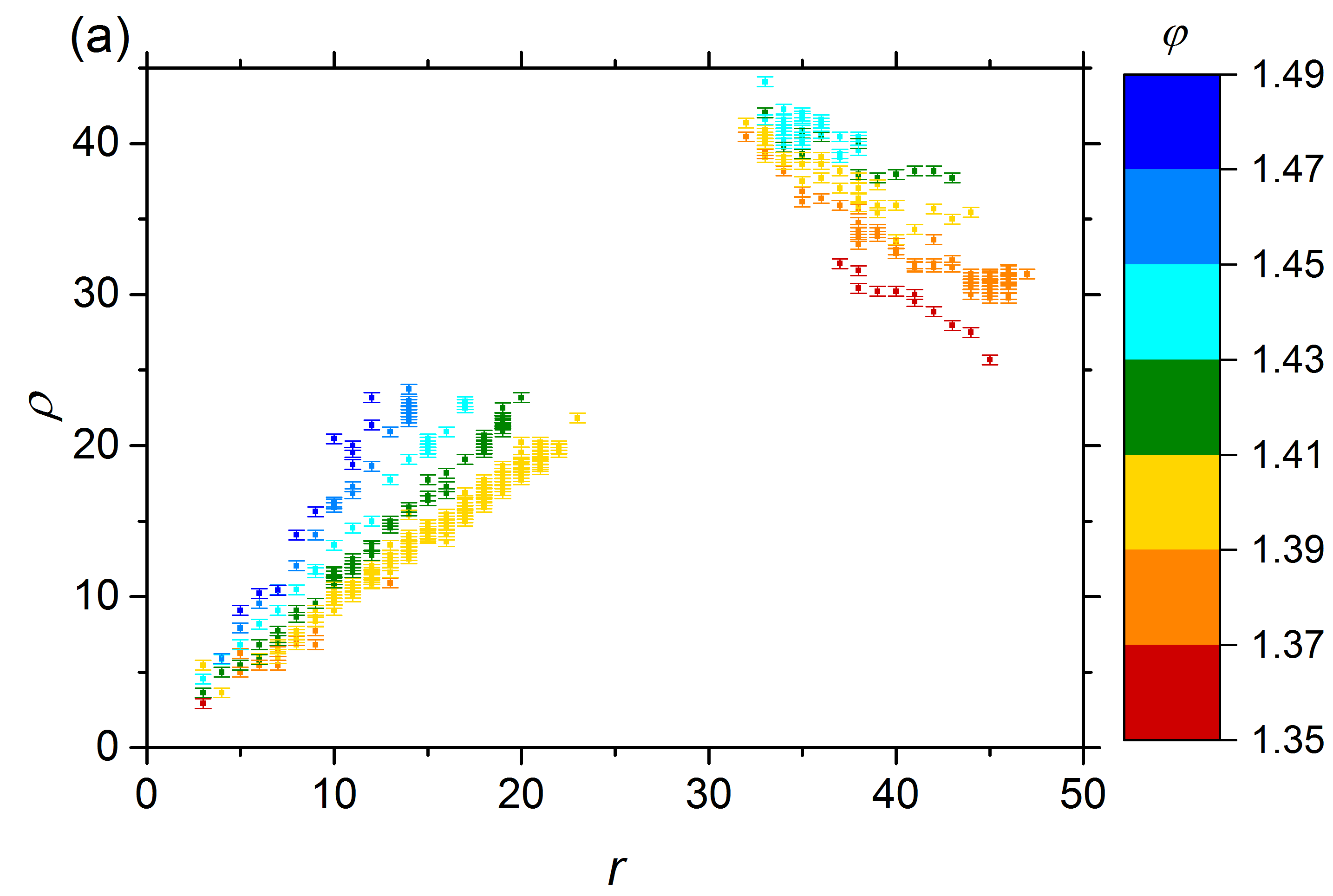}
 \includegraphics[width=1\linewidth]{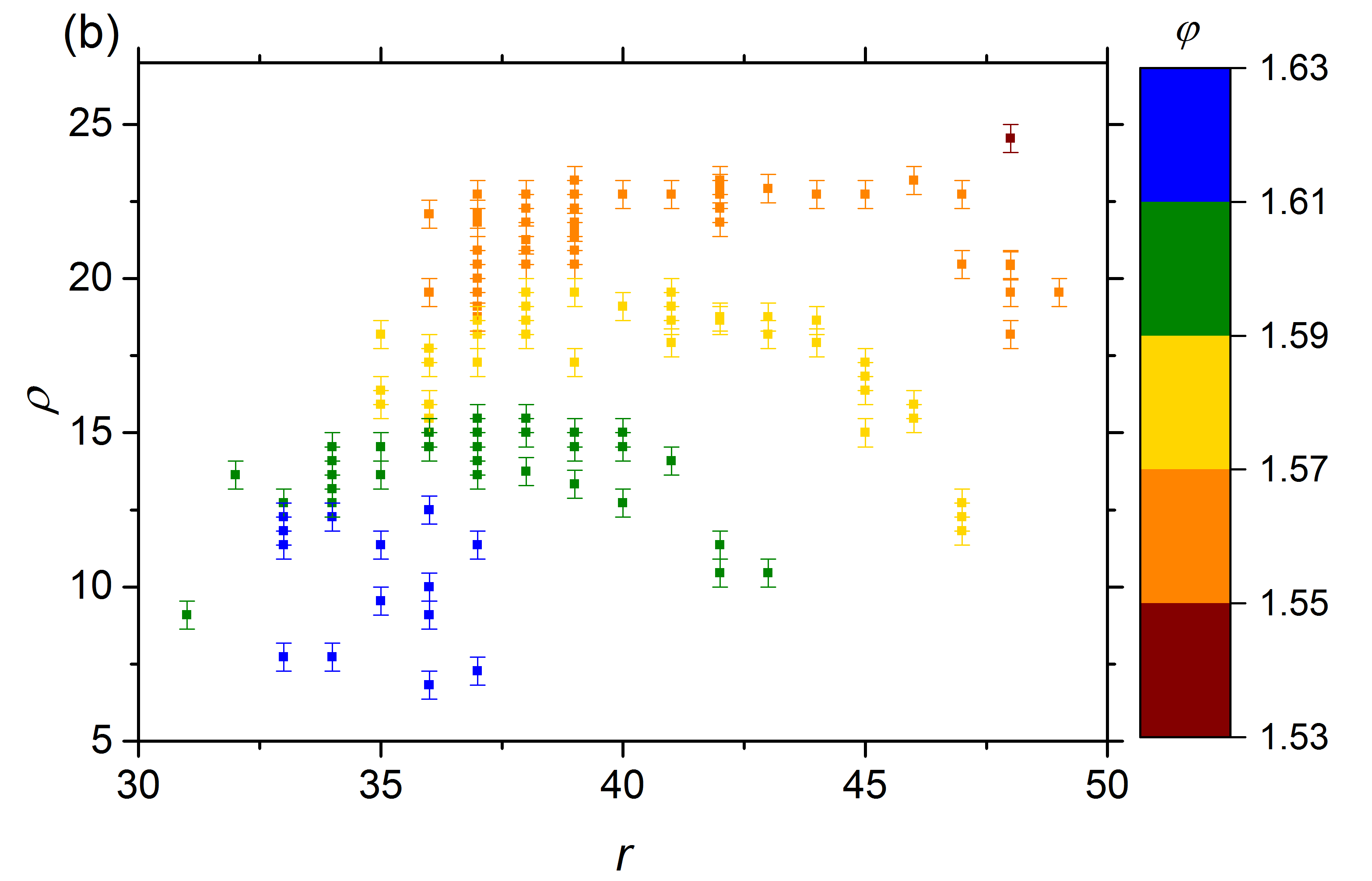}
\caption{(Color online) FHN system: Radius $\rho$ of the spot-like chimera states, (a) incoherent spot and (b) coherent spot, in dependence on the coupling radius~$r$ for different coupling phases~$\varphi$ given as color code. The error bars indicate an uncertainty of (a) $\Delta \rho=0.3$ and (b) $\Delta \rho=0.5$ in the measurements of the radii. Parameters as in Fig.~\ref{fig:fhnSpots1}.}
\label{fig:fhnsizeofspots1}
\end{figure}

An example for a single-headed incoherent spot surrounded by coherent oscillators is shown in the top panel of Fig.~\ref{fig:fhnSpots1}(a). The left, center, and right panels depict a snapshot of $x_{ij}$, the mean phase velocity $\omega_{ij}$, where the average is computed over $1000$ time units, and a horizontal cut of $\omega_{ij}$ through the center of the spot, respectively. The latter exhibits the typical arc-shape profile of a chimera state. When the spots deform to a square shape, the maximum of the $\omega_{ij}$'s is less pronounced and becomes a plateau (not shown). The spots and the squares grow when the coupling radius $r$ increases, as shown in Fig.~\ref{fig:fhnsizeofspots1}(a) for all observed spots. The color code refers to different choices of $\varphi$. We observe two different domains of large and small coupling radii. These correspond to different areas in parameter space, which support spot chimeras, see Appendix~\ref{sec:fhn-parameter-scan}. In the range $20 \leq r \leq 30$ the 
spot patterns lose their stability in space, wander around the grid in an unpredictable way, change their shape over time or even split into pieces and reemerge after some time. 

Another chimera spot pattern -- again found as a single-headed and multi-headed version -- is a spot with a coherent center. An example is shown in the bottom panel of Fig.~\ref{fig:fhnSpots1}(b). As in the previous example, the $\omega_{ij}$-values are lower in the coherent area than in the incoherent region (middle panel). Furthermore, the section of the mean phase velocities 
(right panel) has a shape similar to incoherent spot type and many other chimera 
patterns in one dimension.  This state can also be found with two coherent heads (not shown). The size of the coherent region is depicted in Fig.~\ref{fig:fhnsizeofspots1}(b). We observe only a weak dependence on the coupling radius $r$, while the coupling phase~$\varphi$ has a stronger effect. Compared to the incoherent spot 
chimeras, the size increases with decreasing~$\varphi$. In some cases, we 
observe that at the opposite side of the coherent region of the torus, a spiral can form, see Appendix~\ref{sec:other} for an example. The radius of the spiral increases with the coupling radius (not shown). 

\begin{figure}[ht!]
\centering
 \includegraphics[width=1\linewidth]{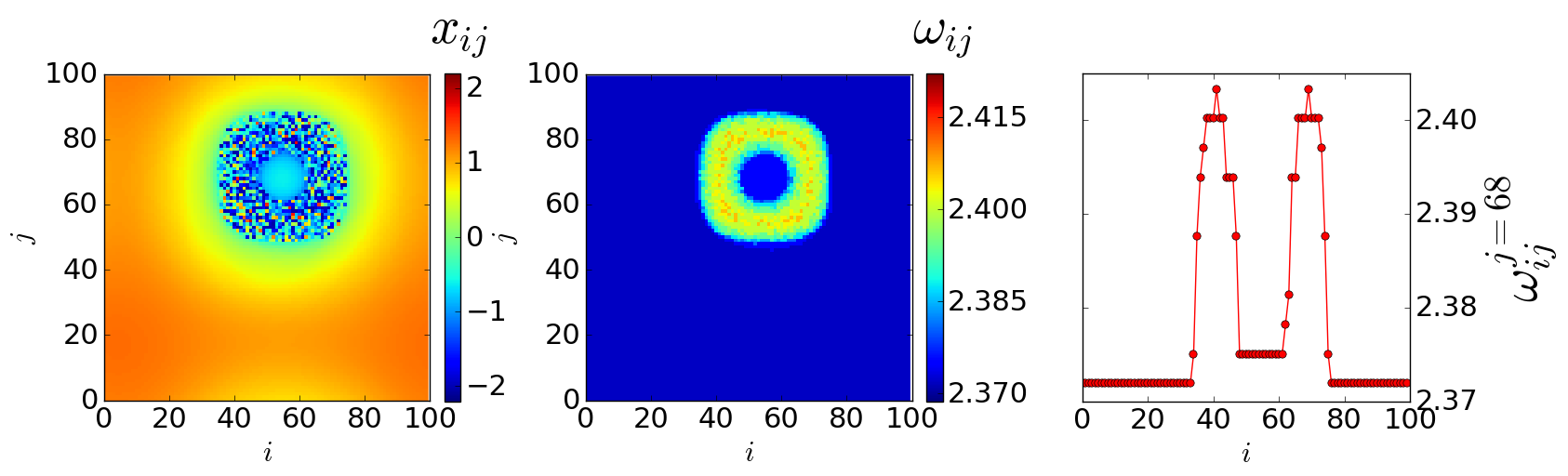}
\caption{(Color online) FHN system: Ring chimera, found for $r=33$ and $\varphi=\pi/2-0.24$. A snapshot at $t=1900$ of the (color coded) activator variable~$x_{ij}$ is shown on the left side and the mean phase velocity and a section through the ring in the center and right panels, respectively. Other parameters as in Fig.~\ref{fig:fhnSpots1}.
}
\label{fig:fhnring}
\end{figure}

\begin{figure}[ht!]
\centering
 \includegraphics[width=1\linewidth]{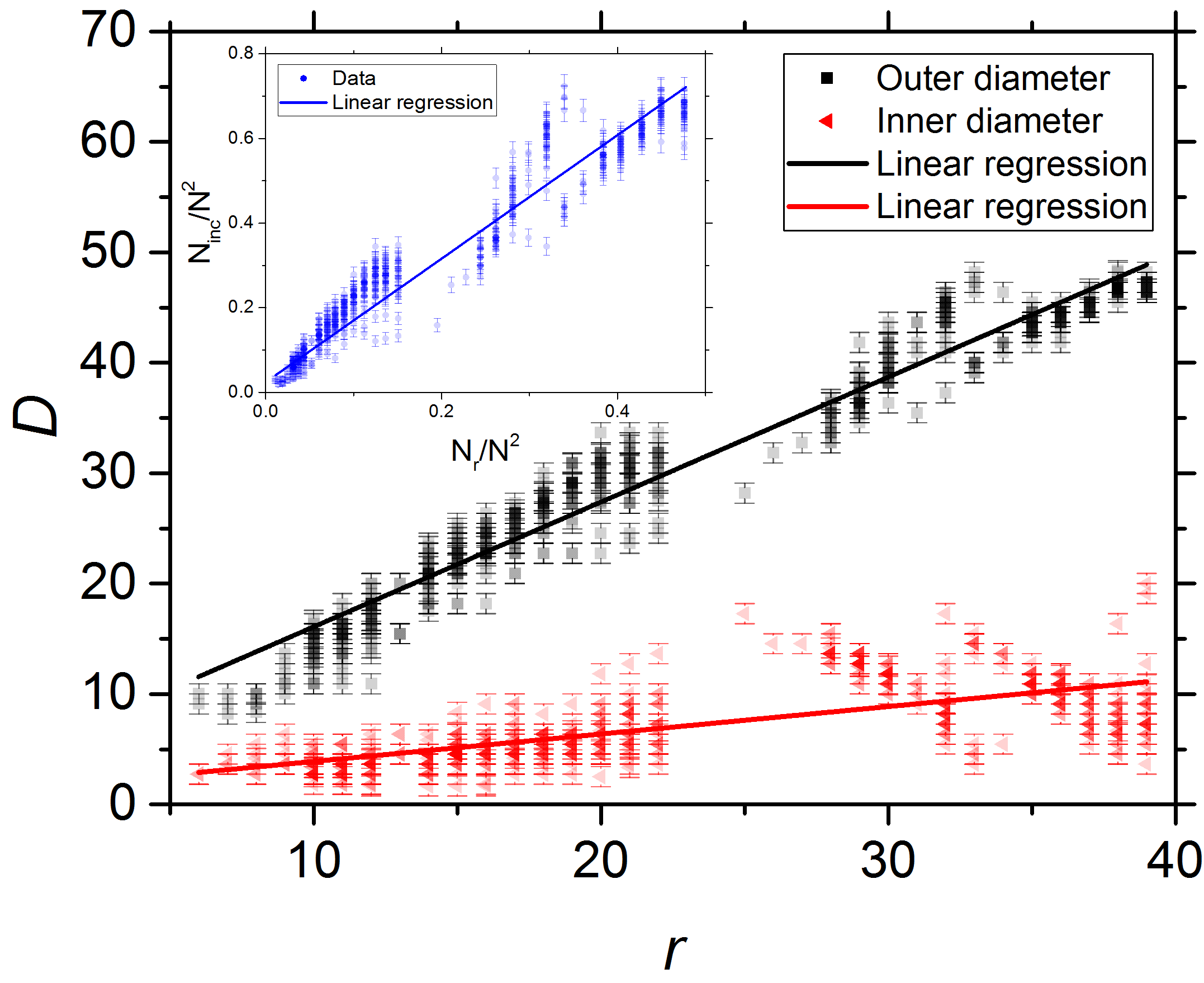}
 \caption{(Color online) FHN system: Inner (red triangles) and outer (black squares) diameter $D$ of the incoherent ring in dependence on the coupling radius $r$. The error bars indicate an uncertainty of $\Delta D=0.9$ in the measurements of the diameters. The lines are linear fits as a guide to the eye. All ring patterns found for $\varphi\in[\pi/2-0.24,\pi/2-0.2]$ are shown. The inset shows the data with rescaled axes: Fraction of incoherent elements $N_{\text{inc}}/N^2$ vs. number of coupled neighbors $N_{r}/N^2$. The brightness of the dots corresponds to the number of observed ring patterns for every ($D,r$)-pair in the considered $\varphi$-range. Other parameters as in Fig.~\ref{fig:fhnSpots1}.
 }
\label{fig:fhn132ringinsideoutside}
\end{figure}

Between regions of spots (see Appendix~\ref{sec:fhn-parameter-scan}) and fully 
synchronous solutions, we observe that ring chimeras can be found for $\varphi <1.35$, see Fig.~\ref{fig:fhnring}.  The label ``ring chimera'' refers to the dip in the mean phase velocity $\omega_{ij}$ in the center of the incoherent ring that can reach the level of the coherent oscillators or have a different, larger $\omega_{ij}$-value as in the displayed example.
We also find that rings can occur as a single-headed state and a multi-headed state with 2 rings 
existing simultaneously at different positions (not shown). 
Similar to the incoherent spots, the rings can either be round or have a square shape. Figure~\ref{fig:fhn132ringinsideoutside} shows the dependence of the inner and outer diameter of the ring chimeras on the coupling radius $r$ using different levels of brightness to indicate the number of occurrences on the considered $\varphi$-range. We find that the outer diameter increases linear with $r$, while the size of the inner ring does not show such a strong dependence. 

The region in parameter space discussed above includes many other patterns, like square- or cross-shaped incoherent spot, stripe (cf. Sec.~\ref{sec:stripes}), alternating and more complex chimeras, many of them in a multistable configuration. 
Examples will be discussed in Appendix~\ref{sec:other}.
Moreover, in the limit $r \to 0$ which corresponds to the continuous reaction-diffusion FHN model, the well-known Turing patterns can be found.
In addition, we find grid-like structures and lines of incoherent spots in a different area of the parameter space (cf. Sec.~\ref{sec:grid}). 

\subsection{Leaky Integrate-and-Fire: Spots and ring patterns}
\label{sect:lif-spots}

Spots and ring patterns are also observed in the LIF model given by Eqs.~\eqref{eq:LIF_07} and~\eqref{eq:LIF_08}, in particular for small values of the coupling strength $\sigma$. The spot
pattern exists for $p_r=0$, while for finite refractory
periods the interior of the spot synchronizes and ring patterns are obtained. As in the case 
of the FHN model we also use a $N \times N=100\times 100$ lattice with toroidal boundary
conditions, while here the control parameters are the coupling radius $R$ and 
the refractory period $p_r$. 

\begin{figure}[ht!]
\includegraphics[clip,width=1\linewidth,angle=0]{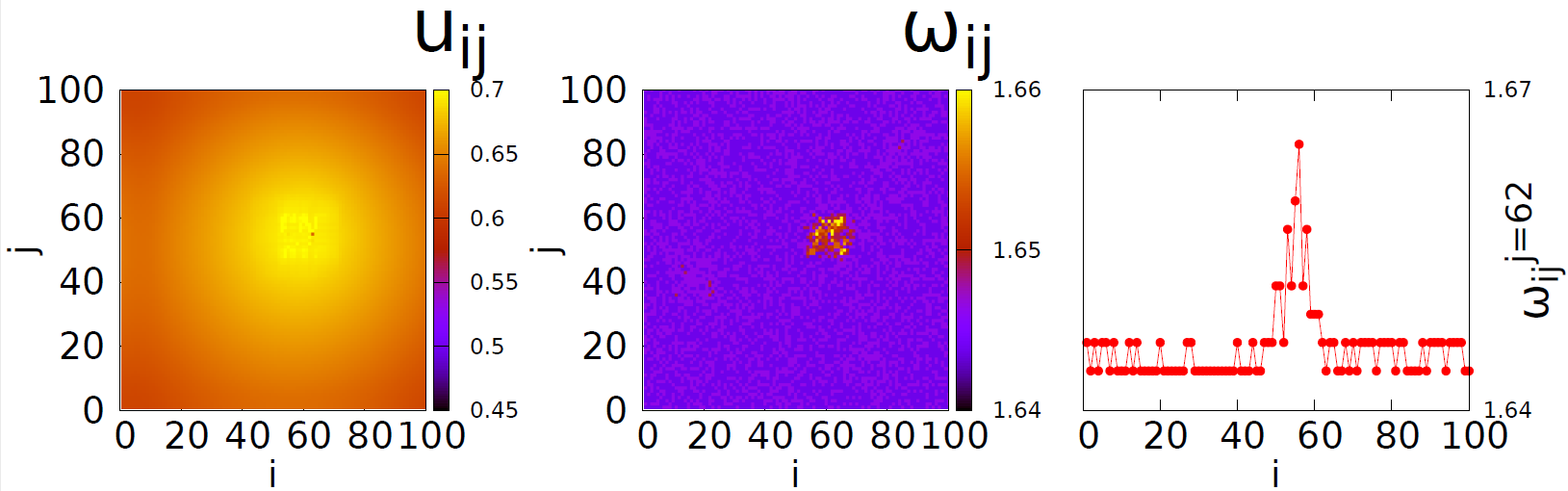}
\caption{\label{fig-lif-spot01} (Color online) LIF system: 
Chimera state of incoherent-spot type. Snapshot of $u_{ij}$ (left) and the corresponding mean phase velocity (middle and right) 
for small values of the coupling constant $\sigma =0.1$ and for a coupling range $R=10$. Other parameters: $\mu =1.0$, $p_r=0$ and
$N=100$.  
}
\end{figure}

Figure~\ref{fig-lif-spot01} depicts an incoherent-spot chimera state for small
values of the coupling constant and zero refractory period. A typical snapshot of the LIF
spot chimera is shown in the left panel, the corresponding mean phase velocity $\omega_{ij}$ is depicted in the middle
and a horizontal cut crossing the spot is depicted on the right. Single spots develop
in the LIF system for small values of $\sigma$ (here $\sigma =0.1$)
when the coupling range $R$ is small. As the coupling range increases while keeping the
refractory period at zero the spot breaks into several unequal 
secondary spots whose size and distance increase with the coupling range (not shown).
%As the size of the coupling range increases and the coverage approaches the system size
%each secondary spot splits further producing asynchronous regions covering homogeneously
%the torus (not shown).

\begin{figure}[ht!]
\includegraphics[clip,width=1\linewidth,angle=0]{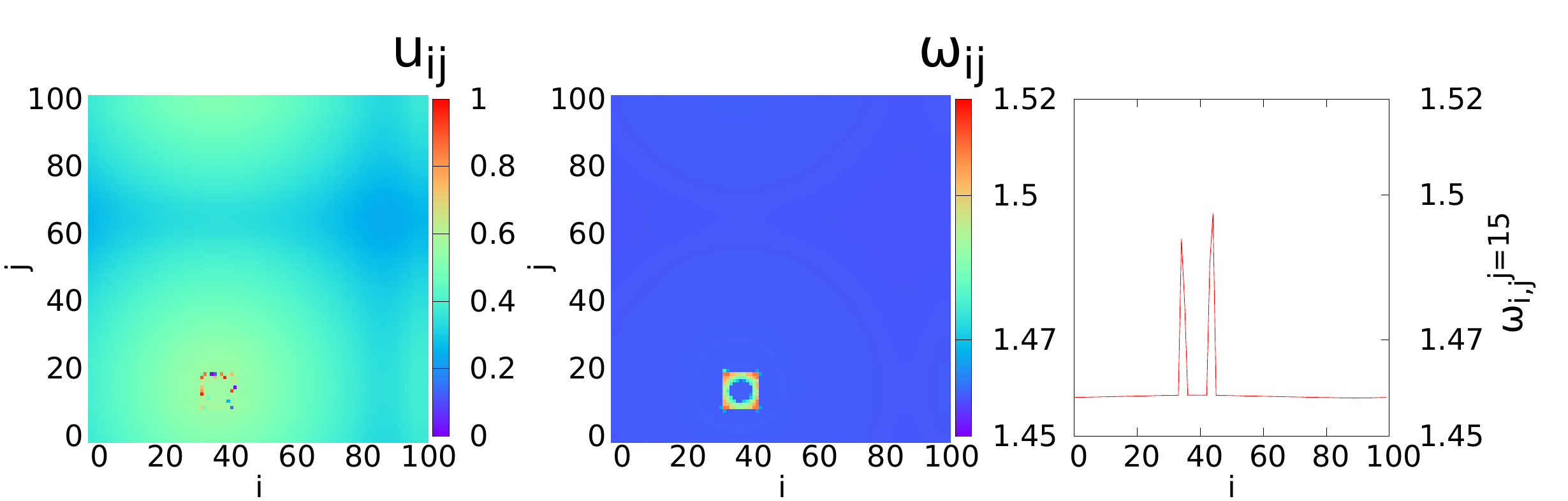}
\caption{\label{fig-lif1} (Color online) LIF system: Ring chimera:
Snapshot of $u_{ij}$ (left) and the corresponding mean phase velocity (middle and right) for a coupling range of $R=10$ and $p_r=0.22 T_s$. Other parameters as in Fig.~\ref{fig-lif-spot01}.
}
\end{figure}

Ring patterns are produced, when the refractory period takes finite values $p_r > 0$ and the interior of the spots synchronizes.
The snapshot of $u_{ij}$, $\omega_{ij}$ and 
its corresponding horizontal cut of a ring chimera is shown in Fig.~\ref{fig-lif1}. 
The outer border of the ring has a square shape due to the square configuration of the connectivity
matrix given by Eq.~\eqref{eq:LIF_08}. This pattern differs from the ones appearing 
in the one-dimensional configuration: From every point of the incoherent 
ring one can reach all other incoherent elements by moving continuously on the ring, and this region forms a set that is not simply connected and cannot exist in one dimension.
As the size of the coupling range increases, the ring radius increases accordingly. 
When the coupling range approaches the
system size, the single ring gives rise to multiple parallel lines of increasing size (not shown).
In the limit of small coupling ranges (e.g. $R=1$), we do not observe the formation of coherent
and incoherent regions. Instead, Turing-like shapes are formed as in the case of the FHN model.

\begin{figure}[ht!]
\includegraphics[clip,width=1\linewidth,angle=0]{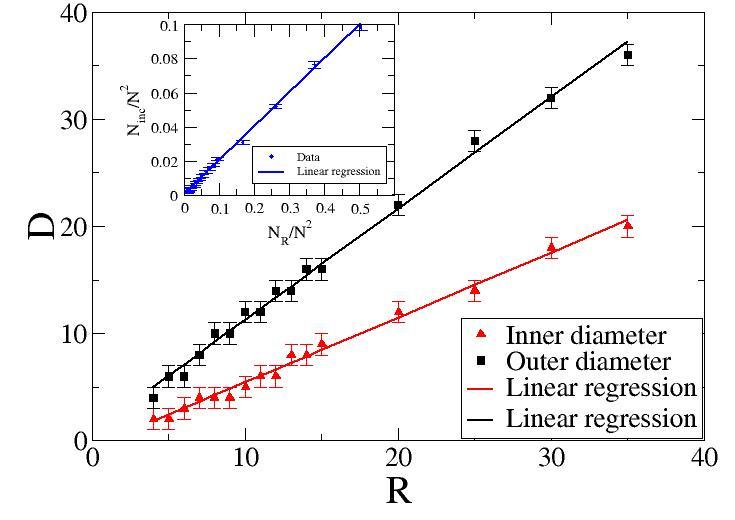}
\caption{\label{fig-lif3} (Color online) LIF system: 
Inner and outer diameter $D$ of the incoherent ring patterns as a function of the coupling range $R$. 
 Inset: Relative number of incoherent elements $N_{\text{inc}}/N^2$ as a function 
of the relative number of coupled neighbors $N_R/N^2$. Parameters as in Fig.~\ref{fig-lif1}. 
}
\end{figure}

The inner and outer diameter
of the incoherent regions as a function of the coupling range $R$ is plotted 
in Fig.~\ref{fig-lif3}. The size of both diameters
increases proportionally to the coupling range as the linear fits 
(solid lines) indicate. 
The inset of the same figure depicts
the relative number of incoherent elements $N_{\text{inc}}/N^2$ (where $N_{\text{inc}}$ is the number of
oscillators in the incoherent domain) as a function of the relative connectivity matrix size $N_R/N^2=(2R+1)^2/N^2$. 
%For the calculation of $N_{\text{inc}}$
%we first find the mean value of $\omega_{ij}-\omega_{\text{min}}=\omega_{\text{dif}}$ given by the equation: 
%$\frac{1}{N_R}\sum_{ij}(\omega_%%{ij}-\omega_{\text{min})}=\overline{\omega_{\text{dif}}} $. 
%Then we count the number of oscillators where $\omega_{\text{dif}}$ 
%is greater than a certain threshold: $\omega_{\text{dif}} >\kappa\,\overline{\omega_{\text{dif}}}$.
%The constant $\kappa$ is chosen so that small changes of $\kappa$ result in the same $N_{\text{inc}}$. 
%The value we use for calculating $N_{\text{inc}}$ in Fig.~\ref{fig-lif3} is $\kappa=4$. 
From this figure we observe that the growth rate of $N_{\text{inc}}/N^2$ is almost linear. We also find, that the mean phase velocities are independent
of the coupling range (not shown). As we will see in the next sections, the mean phase
velocities are mainly influenced by the coupling constant $\sigma$. 

%weg, steht schon da: The solid lines designate linear fits to the data.
%Figure~\ref{fig-lif3} and inset then show that the size of the asynchronous regions is controlled by the coupling range, while we have observed the mean phase velocities are independent of it.

\subsection{Comparison}

Comparing the FHN oscillator and LIF dynamics, we identify similar spatial structures with spot or ring shape. Their size grows linearly with the coupling range. It is worth stressing that the $\omega_{ij}$-values in the incoherent regions are larger than in the coherent ones. 
For small coupling ranges, i.e. the local coupling regime, we find Turing-like behavior. In the LIF model, the oscillators synchronize for large coupling ranges close to all-to-all connectivity, and in the FHN model they synchronize for coupling phases slightly smaller than in the area where spots and rings are observed.

%\clearpage
\section{Stripes}
\label{sec:stripes}
In this section, we focus on a different type of chimera states that exhibit one-dimensional structures extended in the second spatial dimension: stripes.

\subsection{FitzHugh-Nagumo: Stripes}
\label{sec:stripes_FHN}
\begin{figure}[htp]
\centering
 \includegraphics[width=1\linewidth]{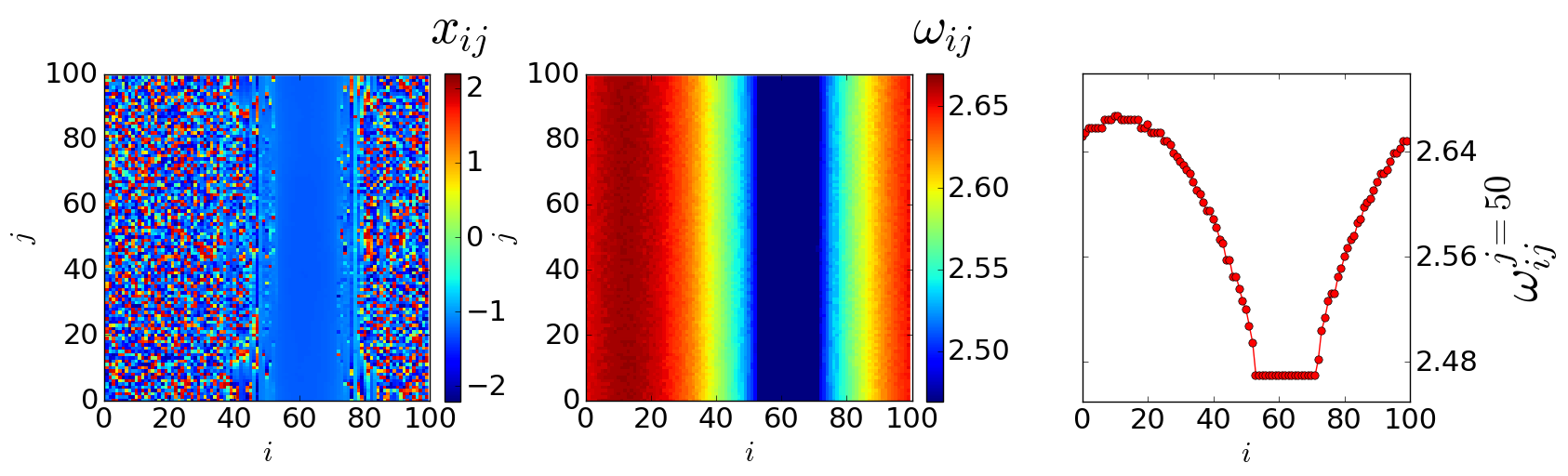}
\caption{(Color online) FHN system: Stripe chimera for $r=39$, $\varphi =\pi/2 -0.04$, and $t=1900$. The (color coded) activator~$x_{ij}$ (left panel) shows a synchronous stripe region, while the other panels show the 
mean phase velocity distribution $\omega_{ij}$ (middle panel) and the section of~$\omega_{ij}$ (right panel). Other parameters as in Fig.~\ref{fig:fhnSpots1}.}
\label{fig:fhnstripe1}
\end{figure}

Besides the two spot types and rings, we also identified stripe chimeras in the FHN model, as shown in Fig.~\ref{fig:fhnstripe1}. These states consist of an incoherent and a coherent stripe region. The latter oscillates with lower mean phase velocity $\omega_{ij}$. Again, the $\omega_{ij}$-profile of the stripe exhibits an arc-shape. In the depicted example the stripe region is synchronous, but in most cases it contains a gradual phase shift, which means that waves travel in the direction of the stripe (see Appendix~\ref{sec:other}). 
For this synchronized stripe type, we never observed a diagonal 
stripe pattern. We also found other stripe patterns that exhibit 
a wave-like dynamics within the stripe instead of complete synchronization. 

\begin{figure}[htp]
 \includegraphics[width=1\linewidth]{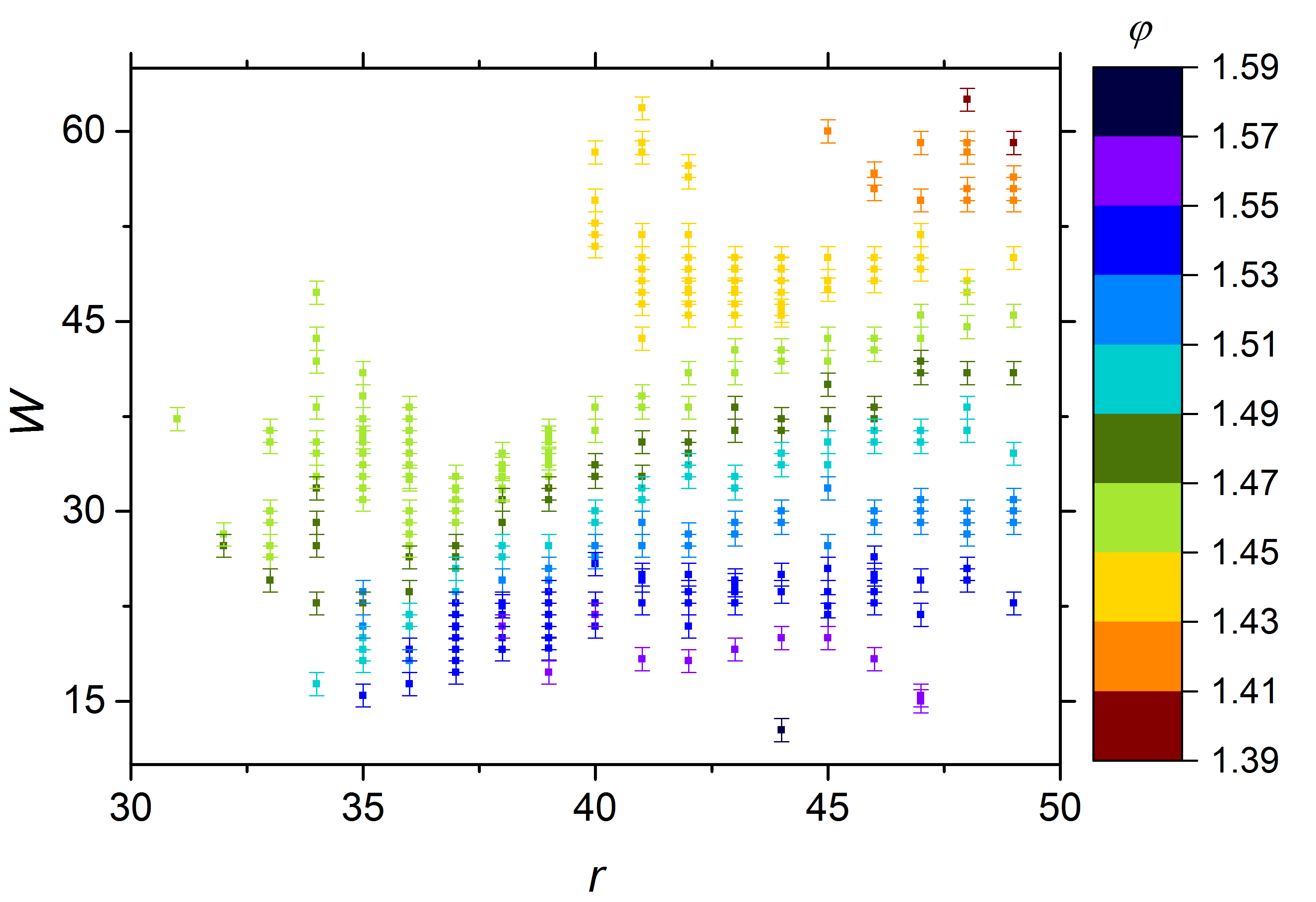}
\caption{(Color online) FHN system: Width~$W$ of the coherent region in the fully synchronized stripes in dependence on the coupling range $r$ for different coupling phases~$\varphi$ as color code. The size of the error bars refers to an uncertainty of $\Delta W=0.9$ in the measurements. Parameters as in Fig.~\ref{fig:fhnSpots1}.}
\label{fig:fhn134stripe}
\end{figure}

Figure~\ref{fig:fhn134stripe} depicts the width $W$ of the coherent region as a function of the coupling range $r$ for the fully synchronized stripe type. The color code refers to different coupling phases. We find that the width of the coherent region increases for larger $r$ and that smaller phases result in larger values of $W$. 
Other multi-stripe chimeras are found outside the previously discussed region of $\varphi\in[1.3,1.65]$ (see Appendix~\ref{sec:other}).

\subsection{Leaky Integrate-and-Fire: Stripes}
\begin{figure}[htp]
 \includegraphics[width=1\linewidth]{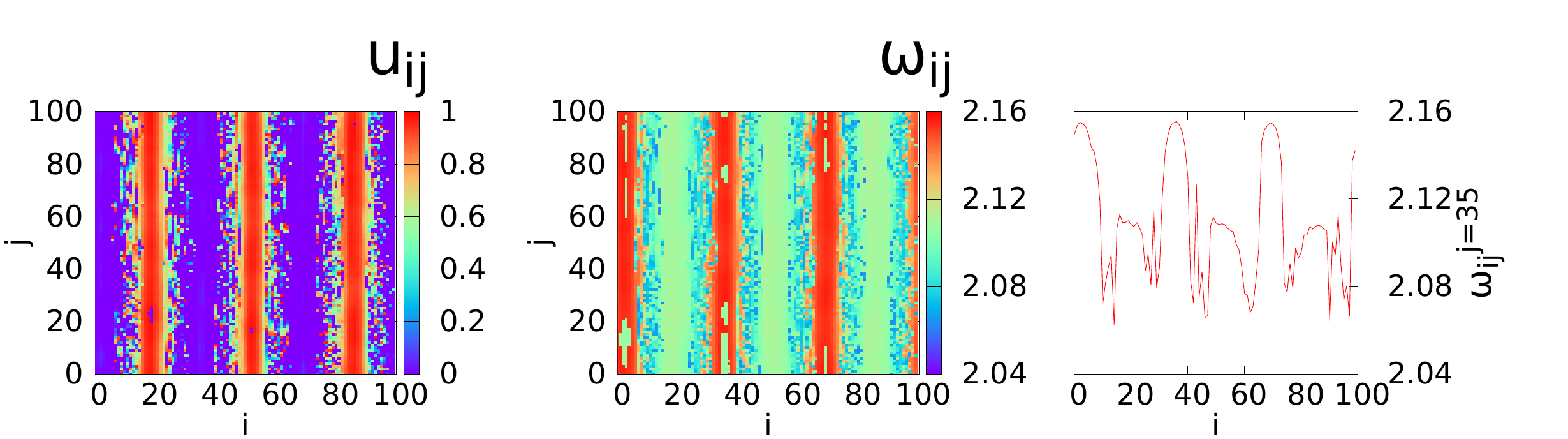}
\caption{(Color online) LIF system: Snapshot of $u_{ij}$ (left), the corresponding mean phase velocity (middle) and a section of it (right). Parameters are $\sigma =0.6$, $\mu =1.0$, $p_r=0.6 T_s$,
$N=100$, and $R=20$. }
\label{fig-lif-stripe01}
\end{figure}

Stripes are also present in the LIF model for large values of the coupling strength and the refractory period, and 
medium values of the coupling range. A stable configuration containing 6 coherent regions
separated by 6 incoherent ones is presented in Fig.~\ref{fig-lif-stripe01}. In fact, the coherent regions
are separated in 2 groups, one with higher mean phase velocity than the other. This is evident from the right panel
of Fig.~\ref{fig-lif-stripe01}, where the section of the mean phase velocity at $j=35$ demonstrates high values
in the order of 2.15 for the first coherent group, intermediate values in the order of 2.10 for the second coherent
group, and $\omega_{ij}$-values between 2.06 and 2.12 for the incoherent stripes. Other stripe multiplicities were also observed
for different values of the coupling range $R$, in the interval $15 \le R \le 25$. For smaller (larger)
values of $R$ a larger (smaller) number of stripes is supported with smaller (larger) width (not shown).
Depending on the initial conditions the stripes form parallel or perpendicular to the $i$-axis. In the LIF model diagonal stripes have not been observed.

\subsection{Comparison}
Stripe chimeras have been found in both FHN and LIF models. They include single and multiple stripes. In the FHN system, we have observed an arc-shaped mean phase velocity profile. The width of the coherent region becomes smaller as the coupling phase increases. For multiple stripes in the LIF model, coherent regions emerge with different mean phase velocities, while $\omega_{ij}$ was found to be equal for every coherent stripe of the multi-headed stripe chimeras in the FHN system. 

\section{Grid Patterns}
\label{sec:grid}

Grids are the most common patterns observed in both FHN and LIF models. These states include features from both spot and stripe chimeras. Examples are presented below.

\subsection{FitzHugh-Nagumo: Grid patterns}
\label{sec:FHN_grid}
The grid chimera is formed by a regular grid of $k\times k$ incoherent spots. 
The spots are either round or spiral-like and have approximately the same size 
given the choice of $r$ and $\varphi$. When approaching the direct-coupling case 
with $\varphi \approx \pi$ they shrink to a size of only a few oscillators and 
disappear. A complex coherent grid pattern without incoherent cores 
remains, which consists of multiple clusters with the same mean phase velocity, 
but non-zero phase lags. 
As seen in the enlargement of the parameter space for the grid chimera states in Appendix~\ref{sec:fhn-parameter-scan}, these cluster states partially 
surround the corresponding grid chimeras. Similar cluster states are also found for the double spot lines and the twisted chimeras discussed below and also in Appendix~\ref{sec:other}. 

Figure~\ref{fig:fhngrid}(a) shows an example of spirals arranged on a grid with asynchronous, incoherent cores.
The grids can also appear as rotated versions with a non-zero angle compared to the underlying lattice structure (not shown). In this work, only the regular grids with the same orientation as the oscillator grid are presented.

\begin{figure}[ht!]
\includegraphics[width=1\linewidth,angle=0]{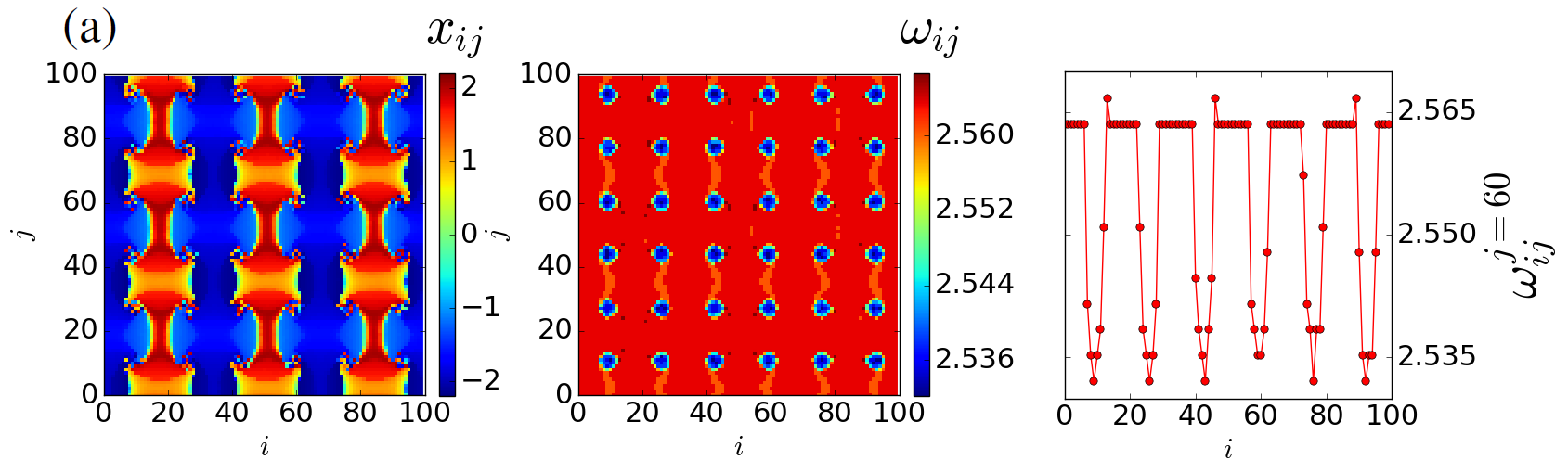}
\includegraphics[width=1\linewidth]{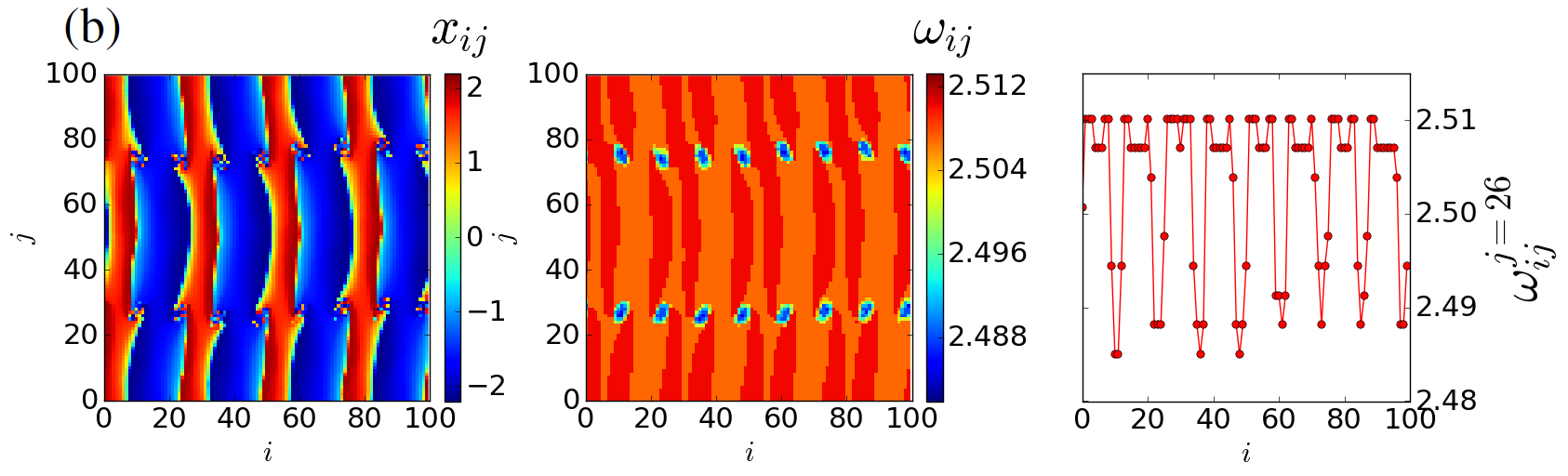}
\caption{(Color online) FHN system: 
(a) Grid state with $6 \times 6$ spots for $r=30$ and $\varphi=\pi/2 +0.82$. The $k \times k$-structure is visible in the (color coded) activator variable~$x_{ij}$ (left) at $t=2000$ and the (color coded) mean phase velocity (center). The right column shows a section of $\omega_{ij}$. (b) Lines of incoherent spots for $r = 22$, $\varphi = \pi/2+0.96$, and $t=1900$. Other parameters in Fig.~\ref{fig:fhnSpots1}.}
\label{fig:fhngrid}
\end{figure}

\begin{figure}[htp]
 \includegraphics[width=1\linewidth]{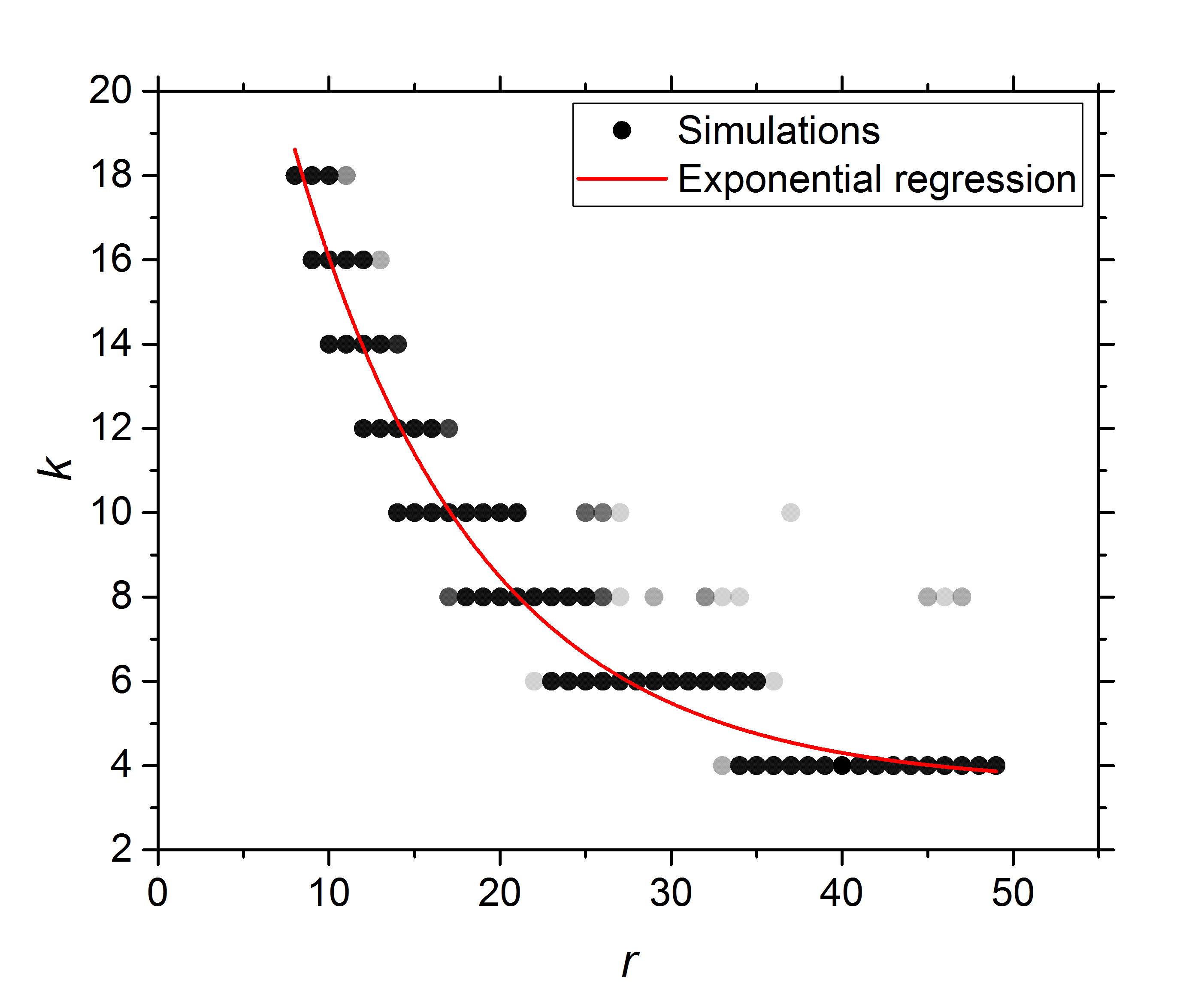}
\caption{(Color online) FHN system: Grid states: Dependence of the number of asynchronous heads $k$ in a single line on the coupling radius~$r$. The red line is an exponential fit. The brightness of the dots indicates the number of occurrences of $k \times k$ grid chimeras for $\varphi \in (1.71,4.65)$. Parameters as in Fig.~\ref{fig:fhnSpots1}.}
\label{fig:gridexp}
\end{figure}

The number of spots in the baseline is found to be even in every grid considered and decreases exponentially with increasing coupling radius $r$, as shown in Fig.~\ref{fig:gridexp}. In case of an odd number of spots in a baseline, the spirals cannot fulfill the boundary conditions. This leads to a destruction of the state. Nevertheless, the grids can exist as a $k\times l$-state, where both $k$ and $l$ are even numbers.

Another grid pattern, which is observed commonly in the two-dimensional FHN model for a large parameter region, is a multi-headed, double-spotted line chimera state. This state consists of two parallel straight lines of incoherent spots, while the surrounding area is filled with a frequency-locked coherent pattern that exhibits a gradual phase shift in space.
An example for such a spotted chimera pattern is shown in Fig.~\ref{fig:fhngrid}(b). 

The spotted patterns can be arranged in a horizontal/vertical or diagonal configuration. When the coupling radius increases, the number of spots in a single line decreases exponentially for both configurations (not shown), similar to the grid chimeras in Fig.~\ref{fig:gridexp}. Again, they can be spiral shaped or round and shrink or disappear completely near the direct-coupling case with $\varphi \approx \pi$, which leaves a coherent pattern. For all coupling phases, every second spot in a line can be shifted off the line for big coupling radii. 

All spotted line states are multistable to other patterns of similar geometry and the grid states. Starting from the basic patterns of two parallel lines of spots, the change of parameters induces a transition to a different chimera pattern, wave patterns, or to synchronization. One possibility is the collapse of the two spotted lines into a single line, which then forms an incoherent stripe. Another possibility is the formation of more than two lines of incoherent spots, mostly 4 lines. Configurations of an odd number of spot lines are observed as well, but they are transients, that is, the lines tend to move, until two of them collide to a single line of spots.

% When we use a parameter set in the $(\varphi,r)$-plane that lies in the center of attraction of one of the spotted patterns, the spot lines are straight and parallel to each other. Using this pattern as an initial condition and exploring the parameter space leads to various scenarios. In short, both lines will start to bend and can either do this in the same direction, against each other or contrary to each other. The chimera state usually dies, if the lines bend against each other, that is, in the direction of their minimum distance. At the contact point, the spots will annihilate, until everything is filled with a wave pattern. If the lines bend in the same direction or contrary to each other, the patterns can exist much longer. 
We also observe that the coherent wave pattern which surrounds the incoherent spots [see Figure~\ref{fig:fhngrid}(b)], can be oriented in space in other ways that include an angle with the underlying grid-structure of the $n \times n$ oscillator grid, while leaving the straight 
lines of incoherent spots intact. We find in a distribution similar to Fig.~\ref{fig:gridexp}, that for a given number of spots $k$ the solutions that are aligned with the grid-structure have big coupling radii, while the radius decreases when the angle increases (not shown).

\subsection{Leaky Integrate-and-Fire: Grid patterns}
\label{sect:lif-grids}

\begin{figure}[ht!]
\includegraphics[clip,width=1\linewidth,angle=0]{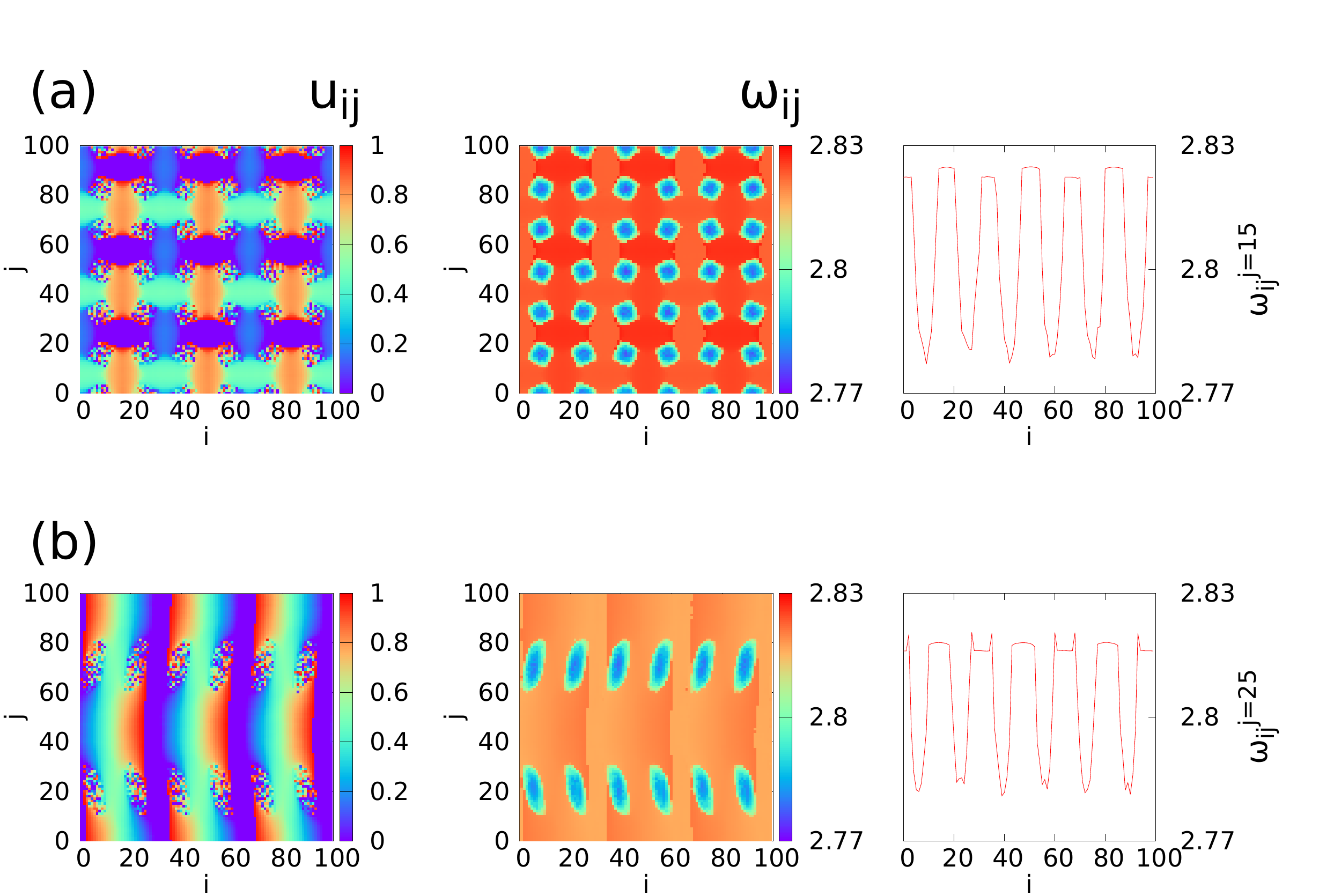}
\caption{\label{fig-lif5} (Color online) LIF system: 
(a) Grid chimera, (b) Double-spotted line. (a) and (b) patterns are formed starting from two different random initial conditions.
Left panels: snapshots of $u_{ij}$; middle panels: corresponding mean phase velocities; right panels: section of mean phase velocities. Parameters: $R=22$, $\sigma =0.7$, $\mu =1.0$, $p_r=0.22 T_s$, and $N=100$. 
}
\end{figure}

\begin{figure}[ht!]
\includegraphics[clip,width=1\linewidth,angle=0]{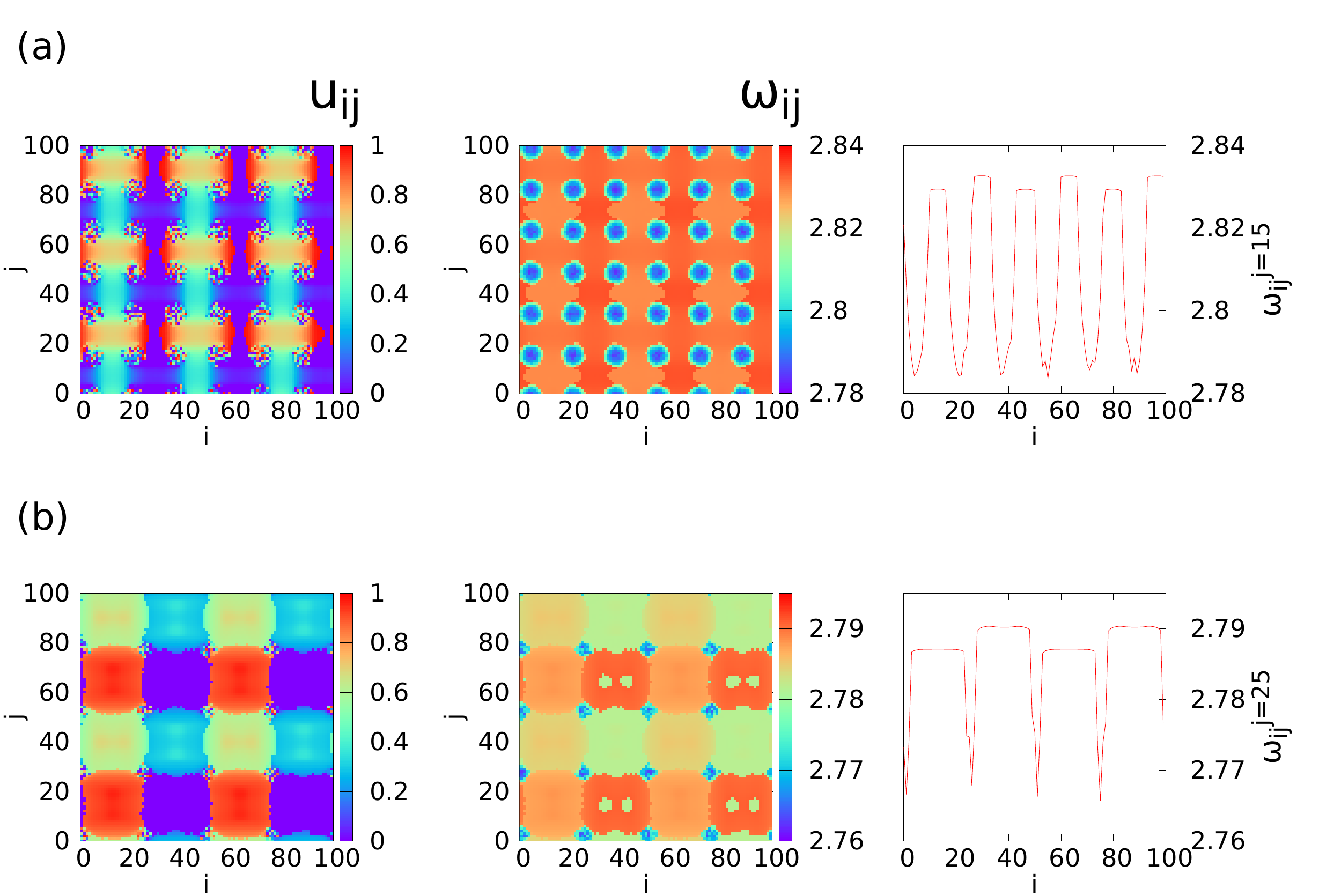}
\caption{\label{fig-lif6} (Color online) LIF system: 
The snapshots of $u_{ij}$ (left panels), 
 corresponding mean phase velocities (middle panels)
and sections of mean phase velocities (right panels) 
 in the limit of strong coupling for two
 different coupling ranges: (a) $R=23$ and (b) $R=45$. 
Parameters: $\sigma =0.7$, $\mu =1.0$, $p_r=0.22 T_s$, and
$N=100$.
}
\end{figure}

In the LIF network, grid chimera patterns emerge when the coupling range takes
relatively high values, see Figs.~\ref{fig-lif5} and~\ref{fig-lif6}. Before studying
the effects of the coupling range $R$ on the form of chimera states, we 
present the case of multistability observed for high values of $\sigma=0.7$ and 
intermediate values of $R=22$. Figure~\ref{fig-lif5} depicts snapshots of $u_{ij}$ (left panels), 
 corresponding mean phase velocities (middle panels)
and section of mean phase velocities (right panels) for LIF realizations 
starting from two different random initial conditions. In panels (a) we observe a 36-headed chimera arranged in a $6\times 6$ grid,
while panels (b) show a 12-headed chimera in the configuration of two horizontal 
 double-spotted bands with 6 incoherent regions each. 
%  This is a typical case of multistability, since the
% system produces two different states starting from two different random initial conditions, while
% all parameters are the same. 
% For this plot, we use the values $\sigma =0.7$ and $R=22$. 
We must stress that pattern (b) is very rare, but stable in time 
and is present only in a few cases around $22 \leq R \leq 24$ and for specific initial conditions. Most initial conditions support
the 36-headed chimeras.
% which are equally stable in time. 
%From now on, we concentrate on the 36-headed chimera and
%investigate the relative size of the coherent and incoherent regions as a function of $R$. 

In Fig.~\ref{fig-lif6} we present the snapshots of $u_{ij}$ (left panels), 
 corresponding mean phase velocities (middle panels)
and sections of mean phase velocities (right panels) 
 for LIF realizations for two different coupling ranges $R$. The 36-headed chimera
persists as we increase the coupling range, up to $R\sim 30$. Above this
value the system synchronizes (not shown). As the limit of all-to-all synchronization is
approached there is another window of partial synchronization around $R=45$,
where a 16-headed grid chimera is observed, see Fig.~\ref{fig-lif6}(b). 

A grouping phenomenon of mean phase velocities is clearly visible in this 16-headed grid chimera pattern
of Fig.~\ref{fig-lif6}(b). The coherent regions split into four groups, with 
each one of them having a
distinct mean phase
velocity, while the $\omega_{ij}$-levels of the
four coherent groups alternate in space. In the right panel of Fig.~\ref{fig-lif6}(b) only two of
these groups are visible since the section for $j=25$ cuts through the 2.785 and 2.79 regions.
Traces of this grouping phenomenon are also visible in Figs.~\ref{fig-lif5}(a)
and~\ref{fig-lif6}(a). 
As the coupling range grows, the difference
between the mean phase velocities of the 
four groups increases  (not shown) and for large values of $R$ the four groups are distinguishable.

Another observation worth studying is that for large values of $\sigma$ the incoherent
regions exhibit small $\omega_{ij}$-values, while the coherent
ones correspond to large $\omega_{ij}$'s. Comparing the
mean phase velocities between Figs.~\ref{fig-lif1} and~\ref{fig-lif6},
we observe that their values have been doubled. 
% As mentioned in sect.~\ref{sect:lif-spots}, the coupling range $R$ does not influence the mean phase velocity. 
We find that it is
the coupling strength that controls the magnitude of them.
As mentioned in Sec.~\ref{sect:lif-spots} for small values
of $\sigma$ the $\omega_{ij}$-values of the incoherent
regions are larger than the coherent ones, while we observe here
that the opposite is true for large values of $\sigma$. This qualitative change
indicates that there is an intermediate value of the coupling constant
where this qualitative change takes place. For the LIF model this critical 
point is located between the values $0.2 < \sigma_{\rm crit} <0.6$. It should
correspond to full synchronization or traveling waves at the cross-over $\omega_{\rm incoh} = \omega_{\rm coh}$. Our simulations show that $\sigma_{\rm crit} \sim 0.3$.
For this value, we observe full synchronization, while for $\sigma$ between approximately 0.4 and 0.5 
the system stays in the transient regime for a long time. This difficulty to attain a
stationary state is another indication that we are near a critical point.

\begin{figure}[ht!]
\includegraphics[clip,width=1.0\linewidth,angle=0]{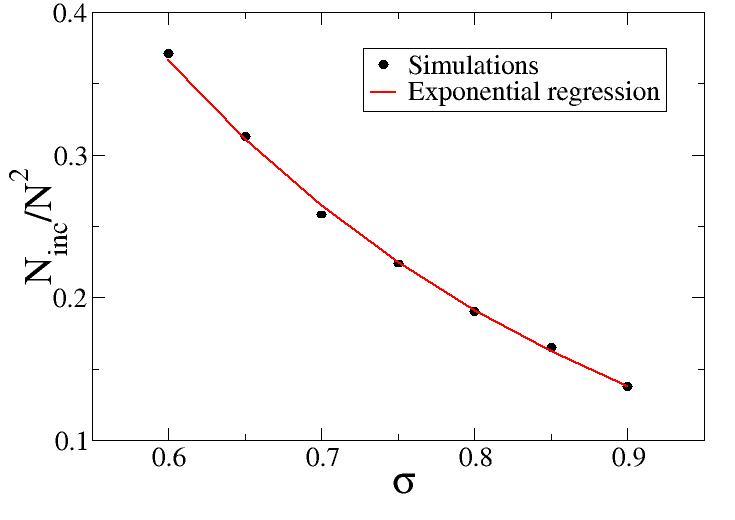}
\caption{\label{fig-lif9} (Color online) LIF system: 
 The ratio of incoherent elements in all observed $6 \times 6$-grids as a function of the coupling strength $\sigma$, using a fixed coupling range at $R =25$ and other parameters as in Fig.~\ref{fig-lif1}.
}
\end{figure}

Overall, the number of incoherent elements decreases with the coupling strength $\sigma$ as shown in Fig.~\ref{fig-lif9}.
This figure is based only on the 36-headed chimera patterns, since the 16-headed is hard to detect in higher $\sigma$ values. The exponential fit (solid line) describes closely the decrease of the size
of the incoherent regions with $R$, with an exponent $\beta =-3.25$ characterizing the decay.
 
The conclusion related to the attributes of the chimera states in the LIF model is that the ratio of coherent and incoherent elements and the mean phase velocities are mainly controlled by the coupling strength $\sigma$. 

\subsection{Comparison} 
Both models support the presence of incoherent spot patterns arranged in grids.
The spots are organized equidistantly along straight lines. The grid consists of an even 
number of such spotted lines in the two spatial directions, along which the number of lines is not
necessarily equal. In the LIF model an odd number of lines is
not observed for the considered parameter scans. In the FHN model spots arranged in odd
numbers of lines may arise, but they soon merge, zipping together and giving rise to 
stable patterns with even number of spots in both directions.

\section{Conclusions and Open Problems}
\label{sec:conclusion}
We have studied the dynamics of two-dimensional networks of neuronal oscillators with nonlocal coupling. 
This can be seen as an intermediate step extending the intensively studied 
one-dimensional ring geometries towards a three-dimensional arrangement of the brain.
FitzHugh-Nagumo (FHN) and Leaky-Integrate-and-Fire (LIF) models have been chosen to represent the single neuronal activity,
as they are two of the most prominent paradigmatic descriptions.
Finding common synchronization patterns in the two dynamical networks could point the way toward identification
of universal dynamical features present in brain activity.

The aim of the current study was to concentrate on stable chimera patterns induced by the nonlocal connectivity and to 
identify patterns that are common in both models. In each model, we have considered a parameter space spanned by three quantities.
For the FHN model the phase connectivity parameter and the coupling range have been varied, while the coupling strength 
was kept fixed. For the LIF model the coupling strength, coupling range, and refractory period were varied.

Our comparative study has demonstrated that, although the dynamics of the single neurons in the two models
are described by different equations, both systems support hybrid states composed of coherent and incoherent regions when the elements are coupled. We have identified a number of common chimera patterns: spots, grids, rings, and stripes. Our simulations suggest that the coherent/incoherent pattern characteristics
follow similar growth rules. For example, the diameter of the
ring patterns grows linearly with the coupling range in both models.
Other phenomena typical in nonlinear systems such as multistability and transitions between
different patterns have been observed as well. The common behavior of the two models supports the universal occurrence of these peculiar dynamics: Chimera patterns are persistent and independent of the specificity of the model, provided that the 
models retain the characteristics of spiky limit cycles.

Future investigations are needed to show whether these chimera patterns will be persistent for a three-dimensional configuration and 
in more complex network architectures, drawing from the current advances in the realistic recordings of
the neuron connectivity in the different parts of the brain. Another aspect worth investigating is the
inhomogeneity of the single neurons. Biological experiments have shown that microscale inhomogeneities 
including different neuron bodies, dendritic structures, axonal fiber bundles etc. are common in the
brain structure and that they affect electrical signal propagation on a microscale~\cite{nelson:2013}. On the
other hand, numerical experiments on the one-dimensional ring architecture support the view that at low levels of
inhomogeneity of the single neuron parameters the chimera patterns persist~\cite{omelchenko:2015}.
It will be insightful to investigate this phenomenon in the case of higher spatial dimensions
with complex connectivity patterns inspired by the ones recorded in brain medical experiments.

\section{Acknowledgments}
AS thanks Yu.~Maistrenko for fruitful discussions. Funding for this study was provided by NINDS R01-40596. 
PH acknowledges support by Deutsche Forschungsgemeinschaft (DFG) in the framework of the Collaborative Research Center 910.
This work was partially supported by the SIEMENS research program on
``Establishing a Multidisciplinary and Effective Innovation and Entrepreneurship Hub''
 and by computational time granted from the Greek Research \& Technology Network (GRNET)
in the National HPC facility - ARIS - under project ID PA002002. 

% \clearpage

% \begin{thebibliography}{10}

% \clearpage
\newpage
\appendix

\section{Parameter scan of the coupled FitzHugh-Nagumo model}
\label{sec:fhn-parameter-scan}
\begin{figure}[ht!]
\centering
 \subfloat[Cartesian distribution of the chimera states.]{\includegraphics[width=\linewidth]{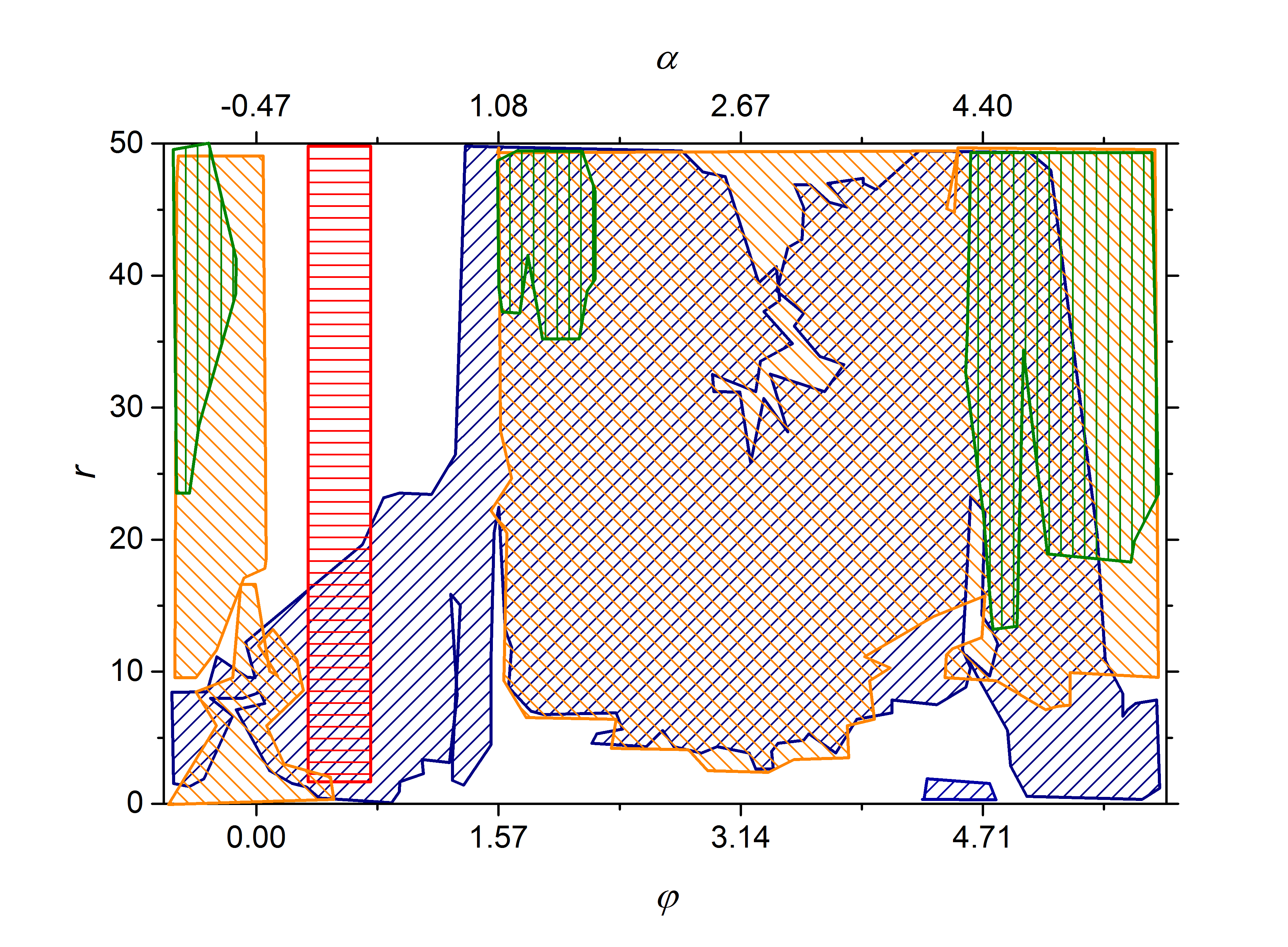}}\\
 \subfloat[Polar distribution of the chimera states.]{\includegraphics[width=\linewidth]{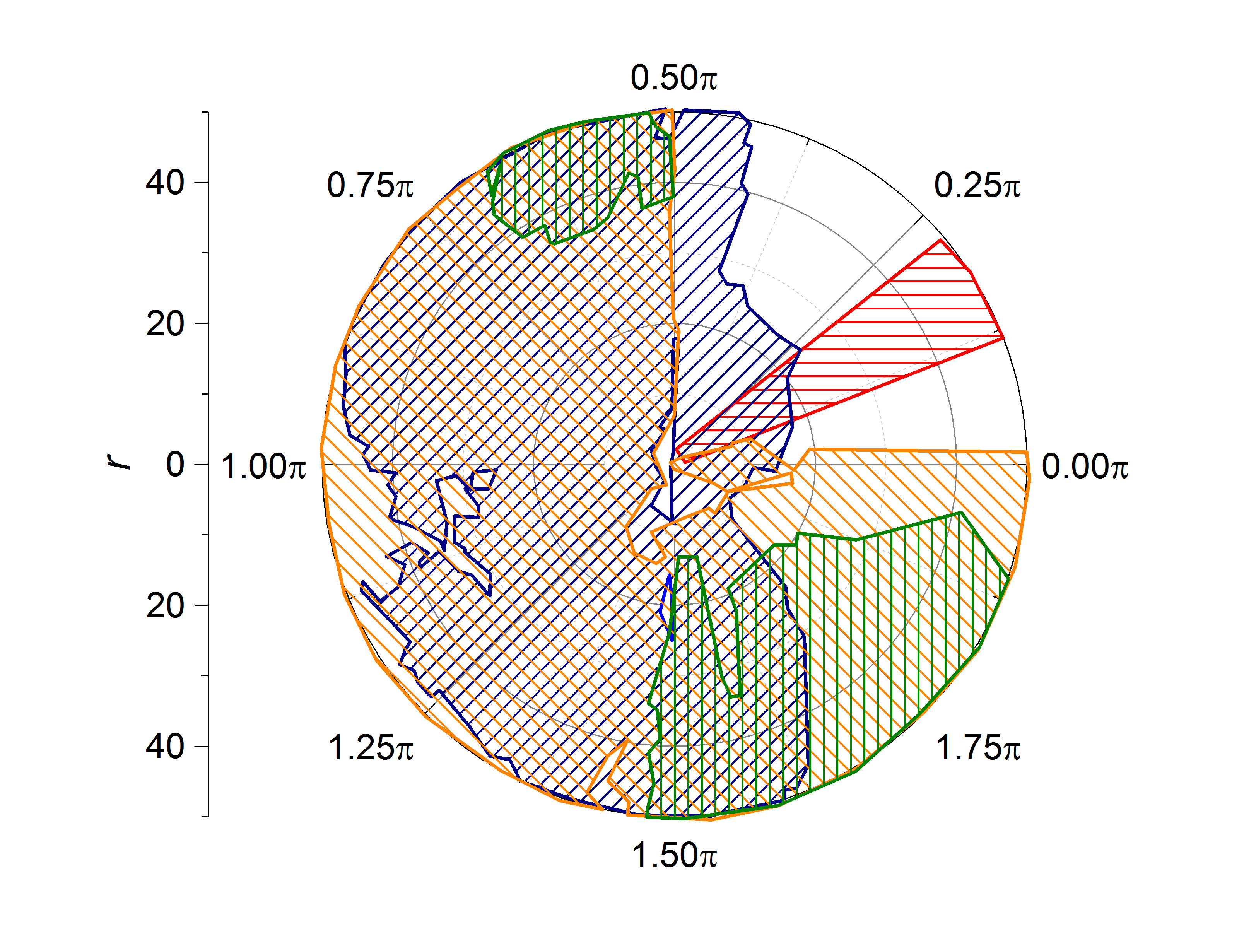}}
\caption{(Color online) FHN system: Full parameter space $(\varphi,r)$. Two different representations of the different states are shown. The colors indicate groups of chimera states (blue), frequency-locked isochronous chimera states (green), cluster states and patterns with a gradual phase shift (orange), and solitary states (red). Other parameters as in Fig.~\ref{fig:fhnSpots1}. }
\label{fig:fhnoverview}
\end{figure}

Figure~\ref{fig:fhnoverview} shows a complete scan of the $(\varphi,r)$-parameter space for the FHN system. The blue region indicates all observed chimera states with a difference in the mean phase velocity $\Delta \omega>0.009$ between the coherent and incoherent region, which means that after $1000$ time units, the oscillators in one of the regions have completed at least $15$ cycles more compared to the other region. The value for $\Delta \omega$ was obtained by visual inspection of the data for different values of $\Delta \omega$.

When the difference of the mean phase velocities is smaller than $\Delta \omega$, the patterns have a very small or no frequency difference between the different regions. These patterns are either fully synchronized or coherent, when a region exhibits a phase lag. The coherent solutions can be cluster states with two or more clusters, which form large synchronous regions, or states with a gradual shift in phase. This type of solutions is colored in orange.

We also find two other coherent cluster solutions that are marked in green and red color. One of them is a two-cluster state shown in green, where the clusters have a phase lag of $\pi$. They are scattered in space and form a spatially chaotic, isochronous region in the oscillator grid, except of one coherent region, where only one of the clusters exists. Such patterns are usually remnant states, evolving from prepared initial conditions (chimeras) and cannot be found with random initial conditions. As these states include two, distinguishable coherent regions, which do not change their position over time, we call them frequency-locked isochronous chimera states.

When one of the clusters is scattered in space in such a way that it only consists of isolated oscillators, which are embedded in the other cluster, the states are called solitary states. These patterns are shown in red.

\begin{figure*}[ht!]
\centering
 \includegraphics[width=1\linewidth]{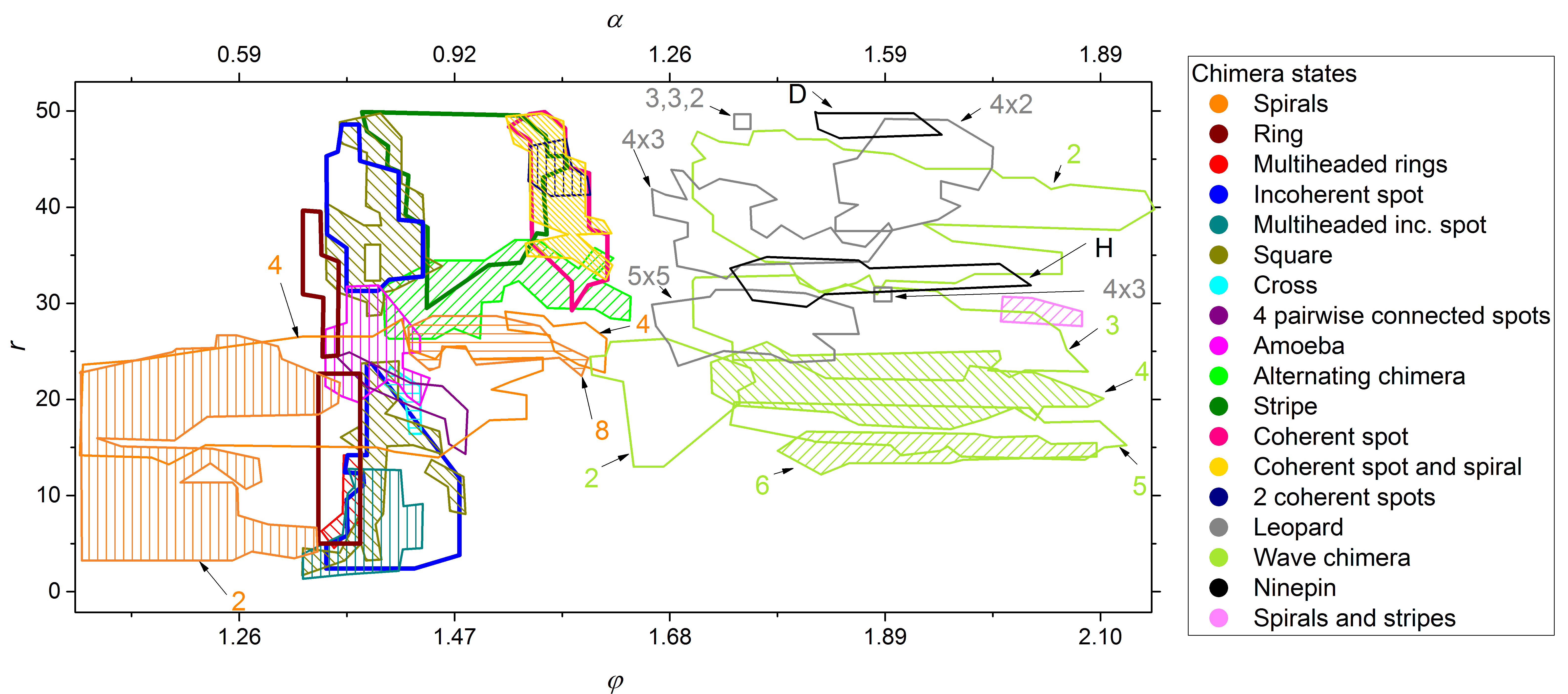}
\caption{(Color online) FHN system: Overview of some chimeras for the cross-coupled region with $1.1 \leq \varphi \leq 2.15$. The regions with bold borders are discussed in the main text in Secs.~\ref{sec:spots_FHN} and~\ref{sec:stripes_FHN}. The colors indicate different types of chimeras, while the arrows and numbers point out the number of coherent heads (for the leopard chimeras), number of spirals (for the spiral states) and the alignment (horizontal, diagonal) of the ninepin chimera. Other parameters as in Fig.~\ref{fig:fhnSpots1}. 
}
\label{fig:fhnoverviewb}
\end{figure*}

Figure~\ref{fig:fhnoverviewb} provides an overview of some of the observed chimera patterns in the range $\varphi\in[1.1,2.15]$. The highlighted areas are discussed in the main text in Secs.~\ref{sec:spots_FHN} and~\ref{sec:stripes_FHN}, some of the other patterns are presented in Appendix~\ref{sec:other}. The numbers in the figure indicate the number of coherent heads (for the leopard chimeras), number of spirals (for the spiral states) and the alignment (horizontal, diagonal) of the ninepin chimera. In this region, many other chimera patterns were also found, like twisted chimeras, grids and double spot lines in different configurations.

\begin{figure*}[ht!]
\centering
 \includegraphics[width=1\linewidth]{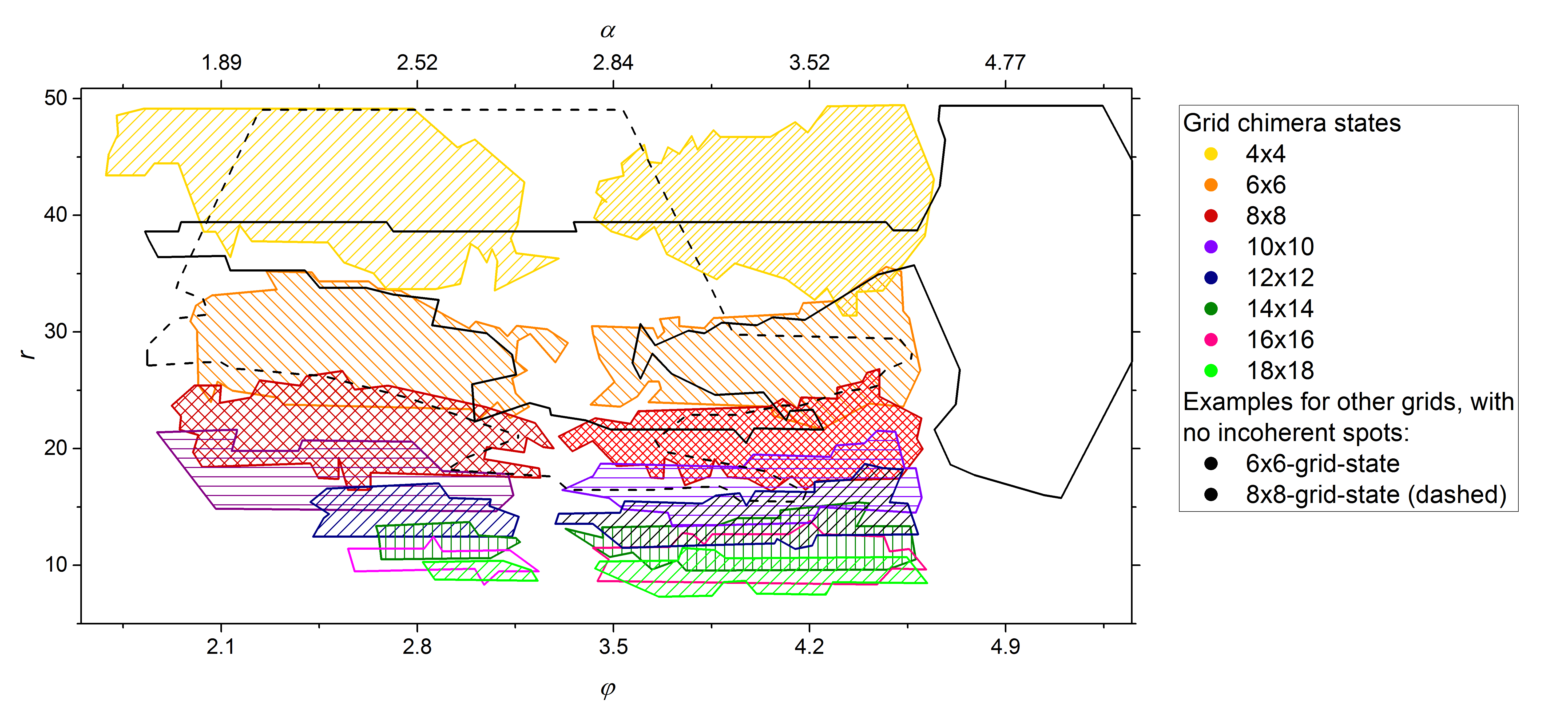}
\caption{(Color online) FHN system: Enlargement of the parameter space where grid chimera states are found (cf. Sec.~\ref{sec:FHN_grid}). The color indicates the number of incoherent heads in a single line. Every grid chimera pattern is partially surrounded by a corresponding grid pattern, which does not include the incoherent cores (black). Other parameters as in Fig.~\ref{fig:fhnSpots1}.
}
\label{fig:overviewcdl}
\end{figure*}

Figure~\ref{fig:overviewcdl} depicts an enlargement of Fig.~\ref{fig:fhnoverview} in the range, where we find grid chimeras. As shown in the main text in Sec.~\ref{sec:FHN_grid}, the number of heads in a single line decreases exponentially with the coupling range $r$. The black lines mark two examples for other grid states that have no incoherent cores. These states are frequency-locked states, which can consist of two or more separately synchronous clusters or regions with a gradual shift in phase.

 \clearpage
\section{Other Patterns}
\label{sec:other}

Apart from the commonly observed spot, stripe, and grid patterns,
a number of other patterns, moving or stable are encountered in each
model. Indeed, we find a plethora of different patterns mainly in the FHN model, 
some of which escape the nomenclature of spots, stripes, or rings:

\begin{itemize}
 \item[(a)] Solitary state, which consists of two clusters with nearly identical mean phase velocities. The number of isolated oscillators grows exponentially with $\varphi$ for every coupling radius $r \geq 7$ on a shared main branch and minor branches for the other radii (not shown). For big coupling radii, the large cluster is synchronous, while for small coupling radii the cluster is coherent and waves reduce the number of solitaries.
 \item[(b)] Double spiral state with static, incoherent cores and a coherent wave pattern which surrounds them. In other examples the cores are moving, often in straight lines, but not necessarily in the same direction. The state can die in the solitary region and form a solitary state, with solitaries arranged in an arc-shape and can also be found with 4 or 8 spirals (not shown).
 \item[(c)] Cross chimera. Although it looks like this is a transitional pattern between a spot and a stripe, they do not exist in the parameter area between these two states.
 \item[d)] Coherent spot and spiral. The spiral grows with coupling range $r$ (not shown). The state also can exist with two coherent spots instead of a spiral.
 \item[(e)] Leopard chimera (multi-headed chimera with $4 \times 3$ shifted coherent heads). The state is also observed with a different number of heads, like $5 \times 5$. The state can be found in a configuration with no mean phase velocity difference between the coherent and incoherent oscillators, when numerical tracking through parameter space is used with prepared initial conditions. Note that the coherent spots have a non-zero phase lag.
 \item[(f)] Ninepin chimera, here in a diagonal configuration. This pattern also exists with a horizontal/vertical alignment.
 \item[(g)] Mosaic chimera. Around the incoherent regions waves are traveling in tubes. Many other stable configurations are also found (not shown).
 \item[(h)] Wave chimera: When taking a vertical cut at $i=60$, $5$ wavefronts can be observed. These states can also exist in other parameter regions with a different number of wavefronts, see Fig.~\ref{fig:fhnoverviewb}. In the example the state exhibits $10$ spots inside the coherent region. In other examples, the chimera does not contain them. 
 \item[(i)] Twisted $5,3$-chimera. Many different twisted chimeras are found. In most cases, the lines are straight, but they can also form curves (not shown) or wriggle. It is possible to create a similar pattern with a different number of wavefronts in the stripe.
 \item[(j)] Organic chimera: No stable pattern is formed in this state, instead it changes its shape continuously. For better visualization, the mean phase velocity is an average over $200$ time units.
 \item[(k)] Diagonal anti-phase stripe and spots. The term \textit{anti-phase} refers to the surrounding pattern, which shares the same mean phase velocity and is coherent, but consists of two or more clusters with non-zero phase lags. Note the two different mean phase velocity levels of the spots and the stripe. The stripe can also exist without the spots and vice versa. The width of the anti-phase stripes and the anti-phase spots grows with coupling range $r$ (not shown), while the number of anti-phase spots decreases exponentially with coupling range $r$.
 \item[(l)] Spiral in an anti-phase spot. Note that there is an incoherent halo around the spot, while the rest of the surrounding pattern is in an anti-phase configuration. The halo-effect was also found for other patterns in the anti-phase region.
\end{itemize}

In the LIF model similar patterns are observed in the simulations, but they were short-lived and transient in most cases.

\begin{figure*}[tp]
\centering
 \subfloat[Solitary state, $r=24$, $\varphi=\pi/2-1.04$ at $t=1900$]{\includegraphics[width=0.5\linewidth]{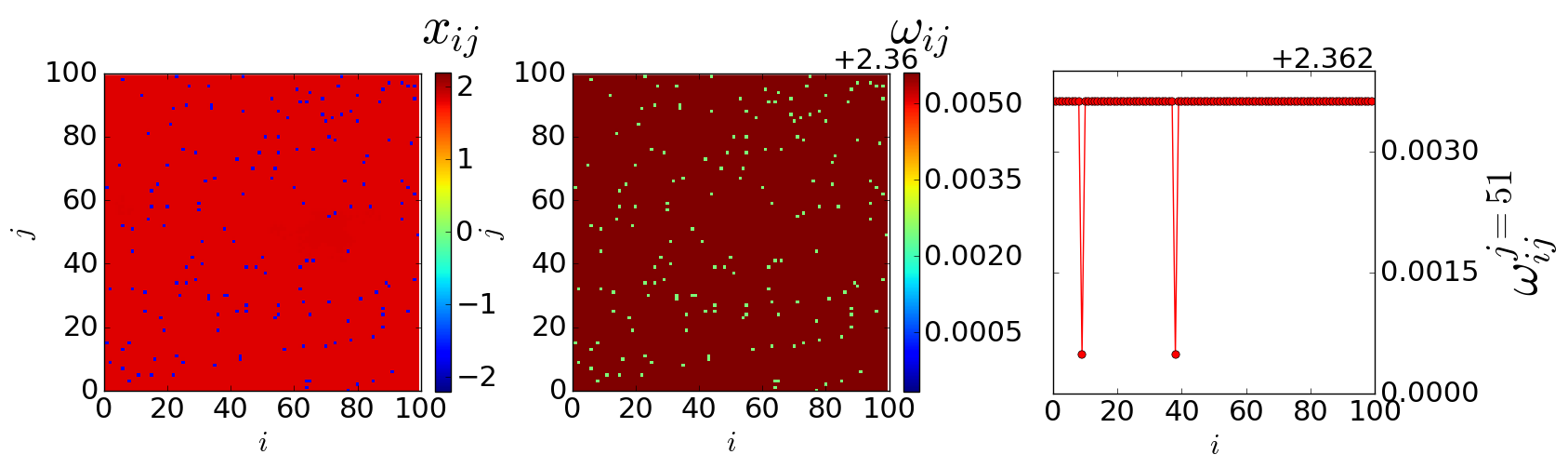}}
 \subfloat[Double spiral chimera, $r=17$, $\varphi=\pi/2-0.54$ at $t=1900$]{\includegraphics[width=0.5\linewidth]{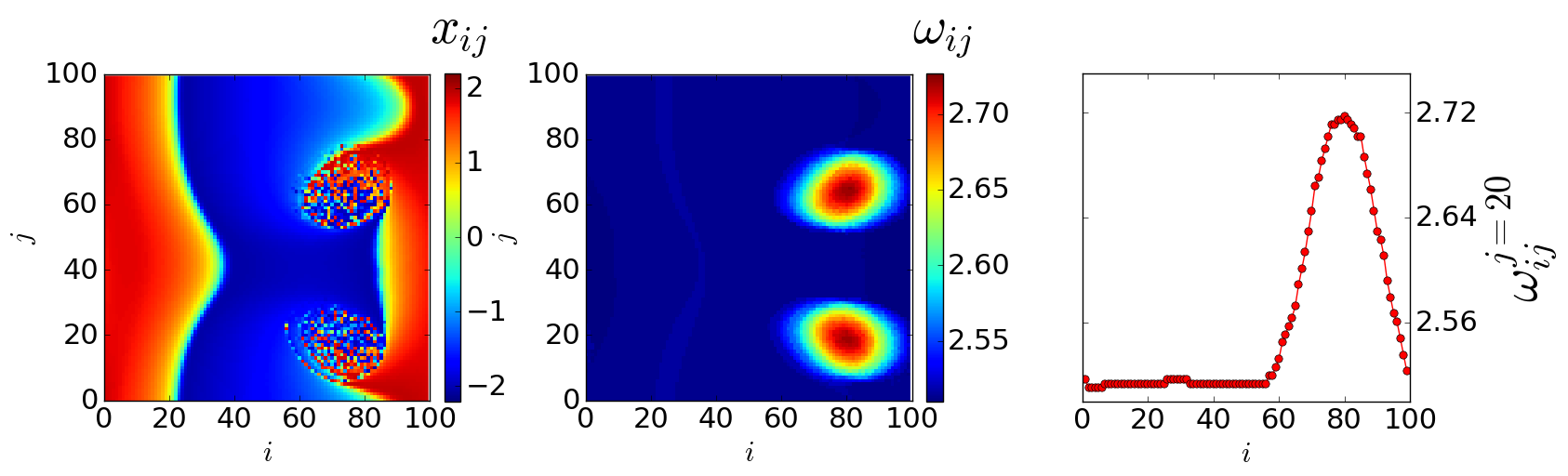}}
\newline
 \subfloat[Cross chimera, 
$r=19$, $\varphi=\pi/2-0.14$ at $t=6000$]{\includegraphics[width=0.5\linewidth]{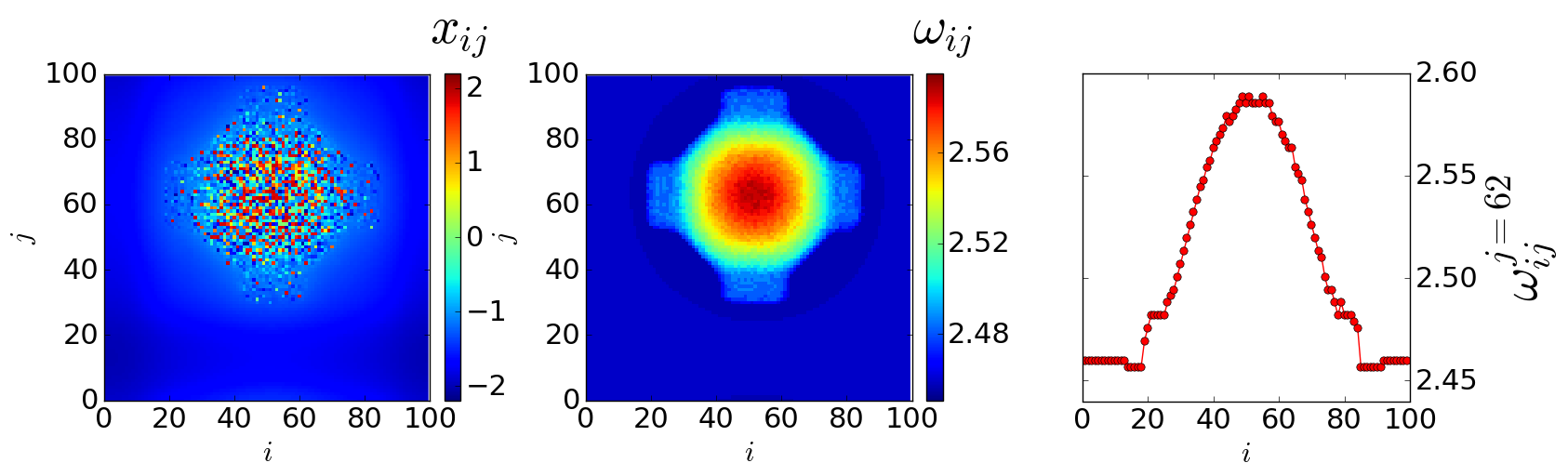}}
 \subfloat[Coherent spot with a spiral, $r=49$, $\varphi=\pi/2-0.02$ at $t=2000$]{\includegraphics[width=0.5\linewidth]{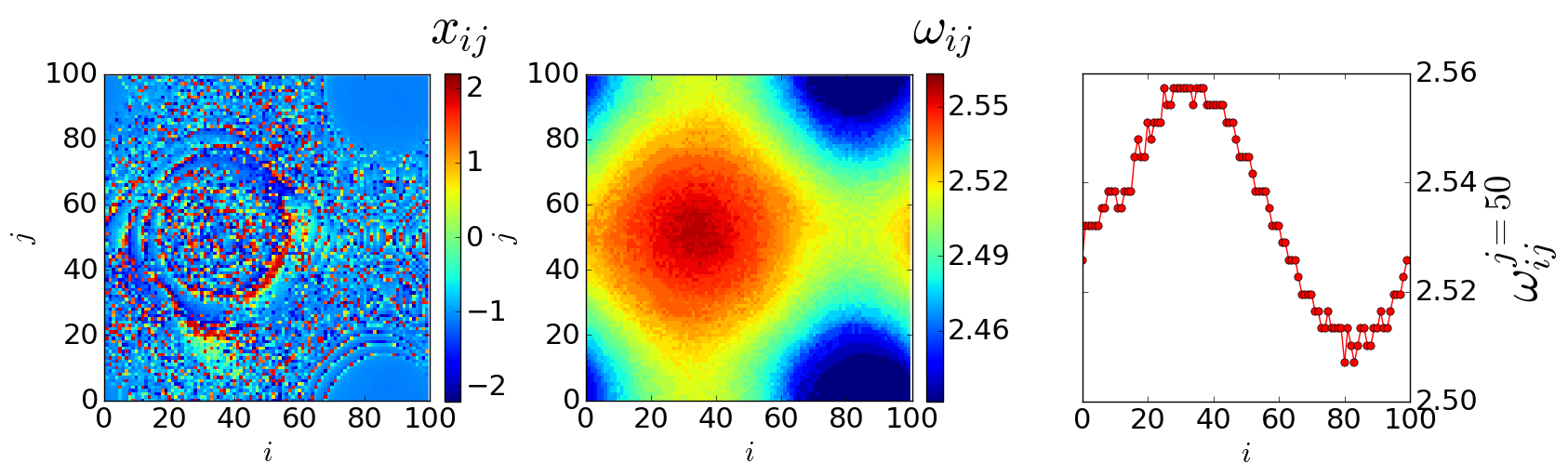}}
\newline
 \subfloat[Leopard chimera ($4\times 3$), $r=37$, $\varphi=\pi/2+0.16$ at $t=2000$]{\includegraphics[width=0.5\linewidth]{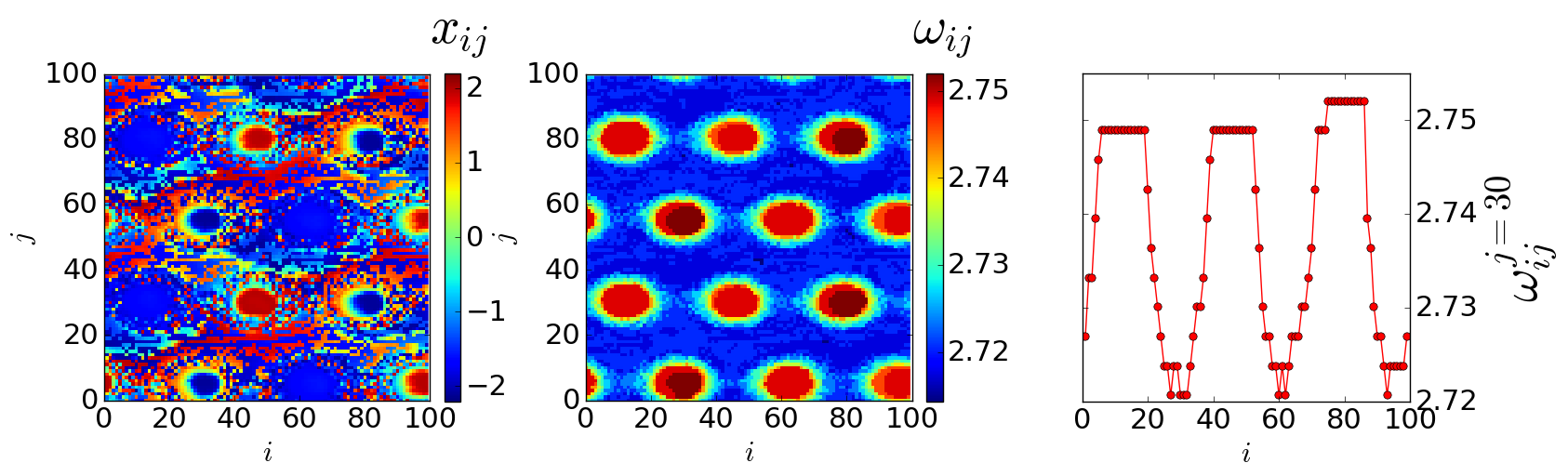}}
 \subfloat[Ninepin chimera ($2\times2$ diagonal), $r=48$, $\varphi=\pi/2+0.34$ at $t=2000$]{\includegraphics[width=0.5\linewidth]{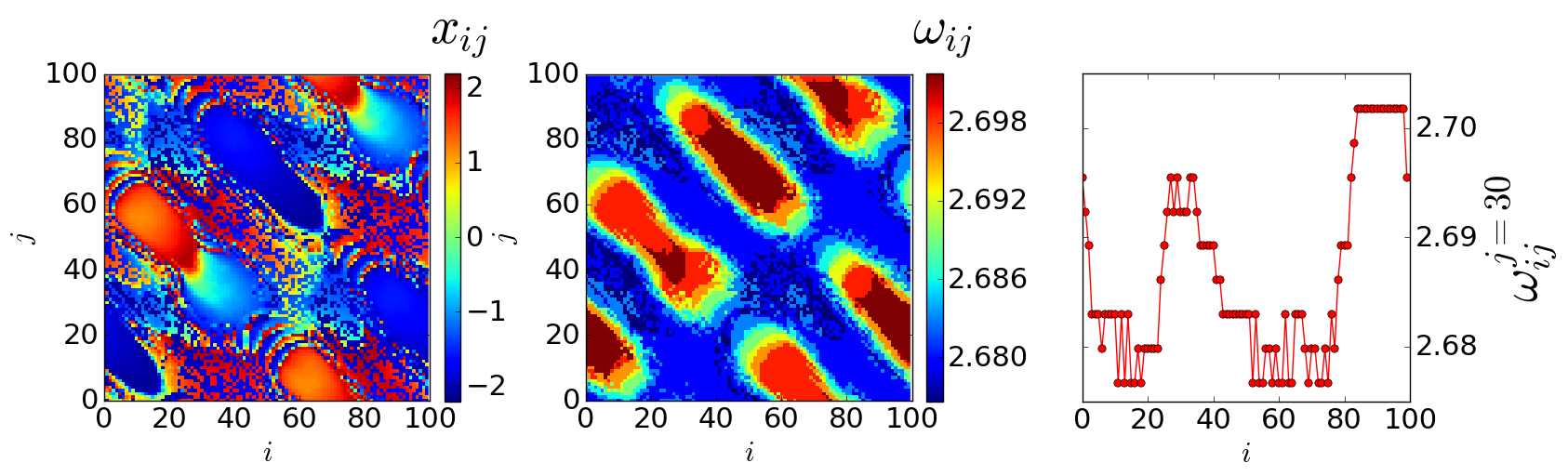}}
\newline
 \subfloat[Mosaic chimera, $r=31$, $\varphi=\pi/2+0.36$ at $t=6000$]{\includegraphics[width=0.5\linewidth]{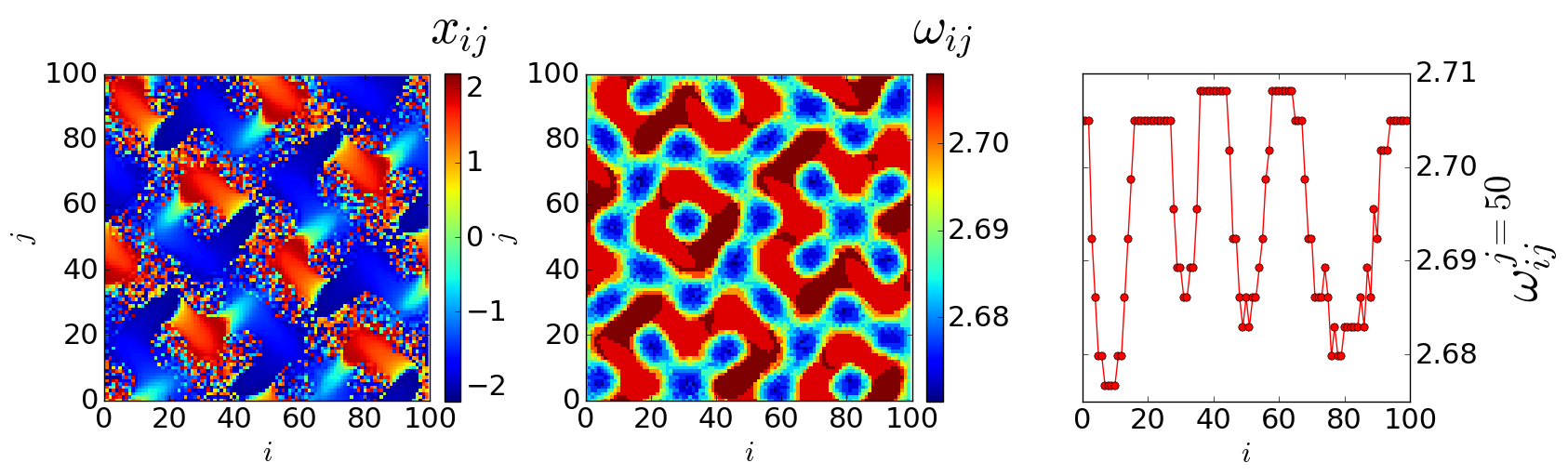}}
 \subfloat[Wave chimera with spots (5 waves), $r=17$, $\varphi=\pi/2+0.48$ at $t=2000$]{\includegraphics[width=0.5\linewidth]{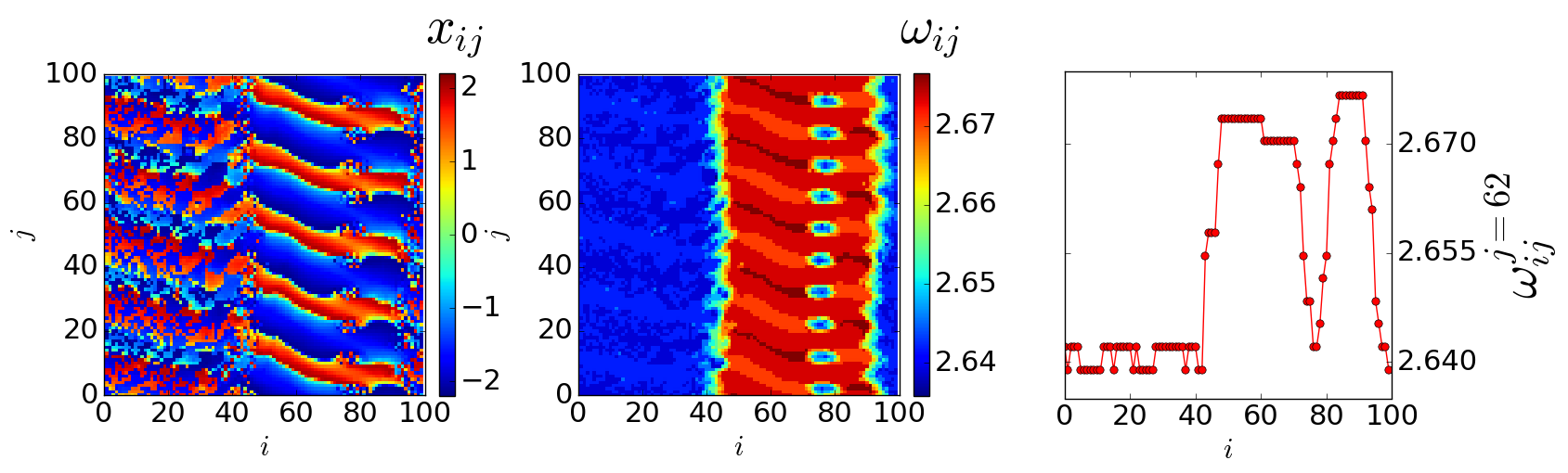}}
\newline
 \subfloat[Twisted 5,3-chimera, $r=25$, $\varphi=\pi/2 +0.42$ at $t=2000$]{\includegraphics[width=0.5\linewidth]{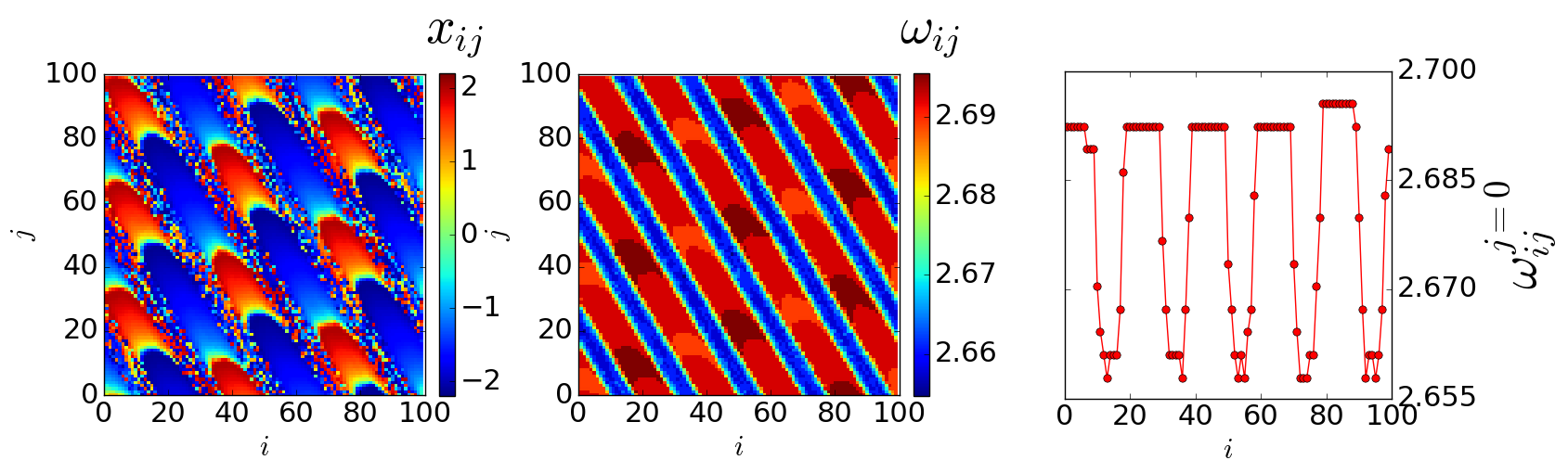}}
 \subfloat[Organic chimera, $r=2$, $\varphi=5\pi/2-2.8$ at $t=6000$]{\includegraphics[width=0.5\linewidth]{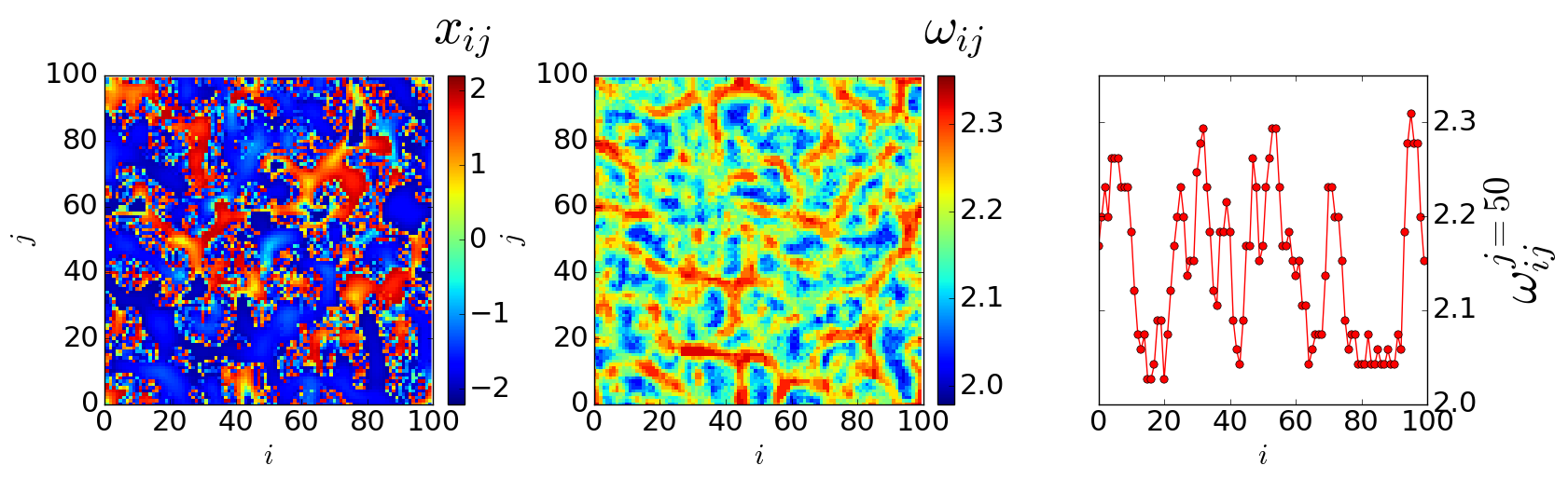}}
\newline
 \subfloat[Diagonal anti-phase stripe and spots, $r=5$, $\varphi=5\pi/2-2.7$ at $t=2000$]{\includegraphics[width=0.5\linewidth]{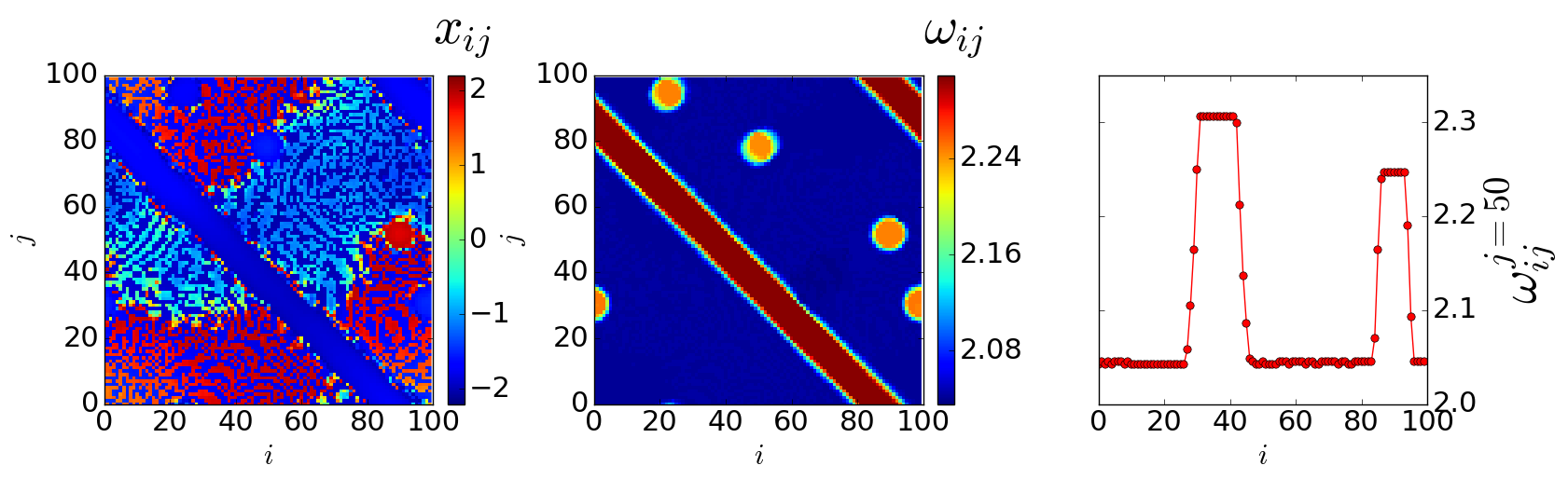}}
 \subfloat[Spiral in anti-phase spot, $r=15$, $\varphi=5\pi/2-2.7$ at $t=1900$]{\includegraphics[width=0.5\linewidth]{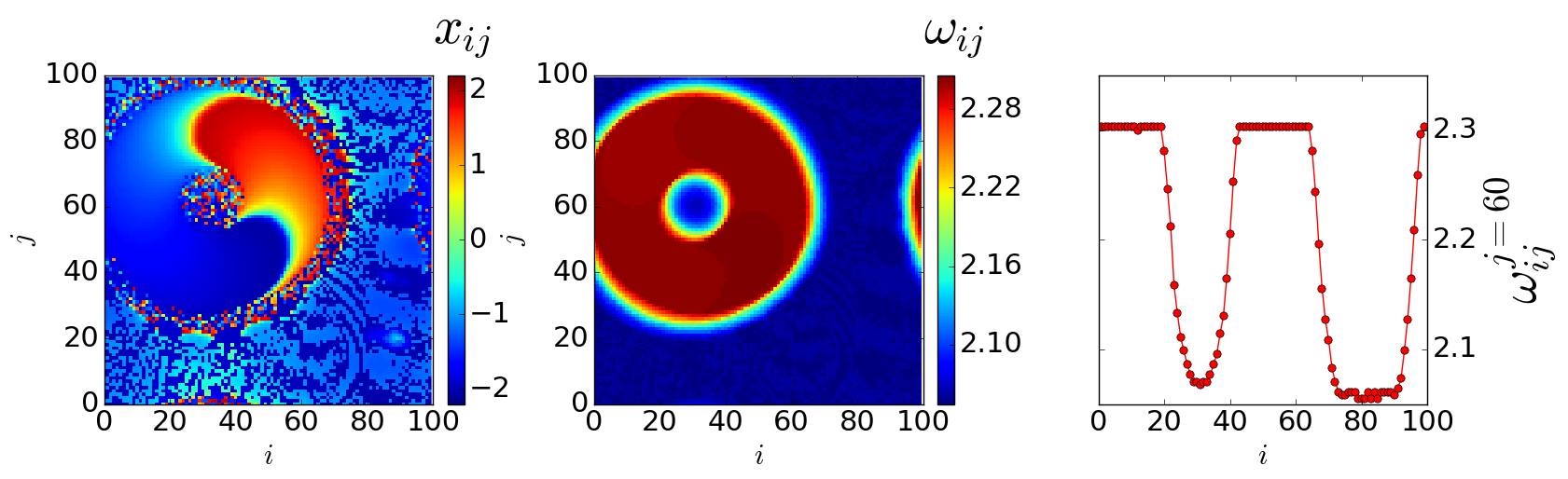}}
\caption{(Color online) FHN system: Examples for chimeras and other states: The left panels show the activator variable $x_{ij}$, while the other panels show the mean phase velocity distribution $\omega_{ij}$ (left panel) and a horizontal section of $\omega_{ij}$ (right panel). Other parameters are $\sigma=0.1$, $a=0.5$, and $\epsilon=0.05$.}
\label{fig:fhnexamples}
\end{figure*}

\end{document}